\documentclass{llncs}
\usepackage{a4wide}

\usepackage{latexsym,amssymb}
\usepackage{graphics,graphicx}
\usepackage{color}
\usepackage[all]{xy}
\usepackage{tikz}
\usepackage{algorithm}
\usepackage{algpseudocode}

\def\R{\mathbb R}

\def\T{\mathbb T}
\def\L{\mathbb L}

         \def\de{\delta} 
\def\ep{\varepsilon}

\def\la{\lambda}        
\def\si{\sigma}

\def\GA{\Gamma}

              \def\bT{\mathbf T}

\newcommand{\SkipIf}{{\bf skipIf}}
\newcommand{\SkipAsn}{{\bf skipAsn}}
\usepackage{ifthen}
\newlength{\upperlabel}
\newlength{\lowerlabel}
\newlength{\translabel}
\newcommand{\trans}{\longrightarrow}
\newcommand{\ptrans}[1]{\Ltrans{#1}{~~~~~~~}}

\newcommand{\pTrans}[1]{\LTrans{#1}{~~~~~~~}}
\newcommand{\LTrans}[2]{%
\settowidth{\upperlabel}{$#1$}%
\settowidth{\lowerlabel}{$#2$}%
\ifthenelse{\lengthtest{\upperlabel>\lowerlabel}}%
{\setlength{\translabel}{\upperlabel}}%
{\setlength{\translabel}{\lowerlabel}}%
\xymatrix@=\translabel{\ar@{=>}[r]^{#1}_{#2}&}}
\newcommand{\Ltrans}[2]{%
\settowidth{\upperlabel}{$#1$}%
\settowidth{\lowerlabel}{$#2$}%
\ifthenelse{\lengthtest{\upperlabel>\lowerlabel}}%
{\setlength{\translabel}{\upperlabel}}%
{\setlength{\translabel}{\lowerlabel}}%
\xymatrix@=\translabel{\ar[r]^{#1}_{#2}&}}
\newcommand{\pLtrans}[3]{%
\settowidth{\upperlabel}{$#1:#2$}%
\settowidth{\lowerlabel}{$#3$}%
\ifthenelse{\lengthtest{\upperlabel>\lowerlabel}}%
{\setlength{\translabel}{\upperlabel}}%
{\setlength{\translabel}{\lowerlabel}}%
\xymatrix@=\translabel{\ar[r]^{#1:#2}_{#3}&}}
\newcommand{\Ntrans}{\xymatrix{\ar@2{)->}[r]&}}
\newcommand{\pNtrans}[1]{
\settowidth{\translabel}{~$#1$~}%
\xymatrix@=\translabel{\ar@2{)->}[r]^{#1}&}}
\newcommand{\Classes}[2]{{#1}/_{#2}}
\newcommand{\Dists}{\mbox{\bf Dist}}

\newcommand{\Confs}{\mbox{\bf Conf}}

\newcommand{\Assume}{\State {\bf Assume:\ }}

\newcommand{\syndef}{\mbox{\tt::=}}
\newcommand{\synalt}{\;\mbox{\tt\large$|$}\;}
\newcommand{\sem}[1]{[\![ #1 ]\!]}
\newcommand{\ttrans}{\Longrightarrow}
\newcommand{\tup}[1]{\langle #1 \rangle}
\newcommand{\false}{\mbox{\bf false}}
\newcommand{\true}{\mbox{\bf true}}

\newcommand{\ifS}[3]{\mbox{\bf if}~#1~\mbox{\bf then}~#2~\mbox{\bf else}~#3}
\newcommand{\whileS}[2]{\mbox{\bf while}~#1~\mbox{\bf do}~#2}

\newcommand{\chooseL}[3]{\mbox{\bf choose}^{#1}~#2~\mbox{\bf or}~#3}
\newcommand{\op}{\mbox{\it op}}
\newcommand{\high}{\mbox{\sc H}}
\newcommand{\low}{\mbox{\sc L}}
\newcommand{\Int}{\mbox{\bf Int}}
\newcommand{\Bool}{\mbox{\bf Bool}}

\begin{document}

\title{
Quantifying Timing Leaks and Cost Optimisation
}
\author{
Alessandra Di Pierro\inst{1} 
\and 
Chris Hankin\inst{2}
\and
Herbert Wiklicky\inst{2}
}
\institute{
University of Verona,
Ca' Vignal 2 - Strada le Grazie 15 
I-37134 Verona, Italy 
\email{dipierro@sci.univr.it}
\and
Imperial College London,  
180 Queen's Gate 
London SW7 2AZ, UK 
\email{\{clh,herbert\}@doc.ic.ac.uk }
}


\maketitle


\begin{abstract}
  We develop a new notion of security against timing attacks where the
  attacker is able to simultaneously observe the execution time of a program
  and the probability of the values of low variables.  We then show how to
  measure the security of a program with respect to this notion via a
  computable estimate of the timing leakage and use this estimate for cost
  optimisation.
\end{abstract}


\section{Introduction}
\label{Introduction}

Early work on language-based security, such as Volpano and Smith's type
systems \cite{VolpanoSmith98b}, precluded the use of high security variables
to affect control flow. Specifically, the conditions in if-commands and
while-commands were restricted to using only low security information. If this
restriction is weakened, it opens up the possibility that high security data
may be leaked through the different timing behaviour of alternative control
paths.  This kind of leakage of information is said to form a {\em covert
  timing channel} and is a serious threat to the security of programs (cf.
e.g. \cite{Kocher96}).

We develop a new notion of security against timing attacks where the attacker
is able to simultaneously observe the execution time of a (probabilistic)
program and the probability of the values of low variables.  This notion is a
non-trivial extension of similar ideas for deterministic programs
\cite{Agat00}\ which also covers attacks based on the combined observation of
time and low variables.  This earlier work presents an approach which, having
identified a covert timing channel, provides a program transformation which
neutralises the channel.

We start by introducing a semantic model of timed probabilistic transition
systems.  Our approach is based on modelling programs essentially as Markov
Chains (MC) where the stochastic behaviour is determined by a joint
distribution on both the values assigned to the program's variables and the
time it takes the program to perform a given command.  This is very different
from other approaches in the area of automata theory which are also dealing
with both time and probability. In this area the timed automata constitute a
well-established model \cite{alur94}.  These automata have been extended with
probability and used in model-checking for the verification of probabilistic
timed temporal logic properties of real-time systems \cite{KNSW04}. The
resulting model is essentially a Markov Decision Process where rewards are
interpreted as time durations and is therefore quite different from our MC
approach. In particular, the presence of non-determinism makes it not very
appropriate as a base of our quantitative analysis aiming at measuring timing
leaks.  We next present a concrete programming language with a timed
probabilistic transition system as its execution model. This language is based
on the language studied in \cite{Agat00}\ but is extended with a probabilistic
choice construct -- whilst this may not play a role in user programs, it has
an essential role in our program transformation.  In order to determine and
quantify the security of systems and the effectiveness of potential
counter-measures against timing attacks we then discuss an approximate notion
of timed bisimilarity and construct an algorithm for computing a quantitative
estimate of the vulnerability of a system against timing attacks; this is
given in terms of the mismatch between the actual transition probabilities and
those of an ideal perfectly confined program.
Finally, we present a probabilistic variation of Agat's padding algorithm
which we use to illustrate -- via an example -- a technique for formally
analysing the trade-off between security costs and protection.


\section{The Model}
\label{Model}

We introduce a general model for the semantics of programs where time and
probability are explicitly introduced in order to keep track of both the
probabilistic evolution of the program/system state and its running time.
 
The scenario we have in mind is that of a multilevel security system and an
attacker who can observe the system looking at the values of its public
variables and the time it takes to perform a given operation or before
terminating, or other similar properties related to its timing behaviour.

In order to keep the model simple, we assume that the time to execute a
statement is constant and that there is no distinction between any `local' and
`global' clocks. In a more realistic model, one has -- of course -- to take
into account also that the execution speed might differ depending on which
other process is running on the same system and/or delays due to
uncontrollable events in the communication infrastructure, i.e. network.

Our reference model is the timed probabilistic transition system we define
below.  The intuitive idea is that of a probabilistic transition system
(similar to those defined in all generality in \cite{JonssonEtAl01}) where
transition probabilities are defined by a joint distribution of two random
variables representing the variable updates and time, respectively.


Let us consider a finite set $X$, and let $\Dists(X)$ denote the set of all
{\em probability distributions} on $X$, that is the set of all functions $\pi:
X \rightarrow [0,1]$, such that $\sum_{x\in X} \pi(x) = 1$. We often represent
these functions as sets of tuples $\{\tup{x,\pi(x)}\}_{x\in X}$.
If the set $X$ is presented as a Cartesian product, i.e.  $X = X_1 \times
X_2$, then we refer to a distribution on $X$ also as a {\em joint
  distribution} on $X_1$ and $X_2$.  A joint distribution associates to each
pair $(x_1,x_2)$, with $x_1\in X_1, x_2\in X_2$ the probability
$\pi(x_1,x_2)$.
It is important to point out that, in general, it is not possible to define
any joint distribution on $X_1 \times X_2$ as a `product' of distributions on
$X_1$ and $X_2$, i.e. for a given joint distribution $\pi$ on $X = X_1 \times
X_2$ it is, in general, not possible to find distributions $\pi_1$ and $\pi_2$
on $X_1$ and $X_2$ such that for all $(x_1,x_2) \in X_1 \times X_2$ we have
$\pi(x_1,x_2) = \pi_1(x_1)\pi_2(x_2)$. In the special cases where a joint
distribution $\pi$ can be expressed in this way, as a `product', we say that
the distributions $\pi_1$ and $\pi_2$ are {\em independent}
(cf. e.g. \cite{Stirzaker99}).


\subsection{Timed Probabilistic Transition Systems}

The execution model of programs which we will use in the following is that of
a labelled transition system; more precisely, we will consider probabilistic
transition systems (PTS). We will put labels on transitions as well as states;
the former will have ``times'' associated with them while the latter will be
labelled by uninterpreted entities which are intended to represent the values
of (low security) variables, i.e. the computational state during the execution
of a program. We will not specify what kind of ``time labels'' we use -- e.g.
whether we have a discrete or continuous time model -- we just assume that
time labels are taken from a finite set $\T\subseteq\R^+$ of positive real
numbers. The ``state labels'' will be taken from an abstract set which we
denote by $\L$.
 
\begin{definition}\label{tPTS}
  We define a {\em timed Probabilistic Transition System with labelled
    states}, or tPTS, as a triple $(S,\trans,\la)$, with $S$ a finite set of
  {\em states}, $\trans \;\subseteq S \times \T \times [0,1] \times S$ a
  probabilistic transition relation, and $\la: S \rightarrow \L$ a state
  labelling function.
\end{definition}

We denote by $s_1 \ptrans{p:t} s_2$ the fact that $(s_1,p,t,s_2) \in \trans$
with $s_1,s_2\in S$, $p\in[0,1]$ and $t\in\T$. In a general tPTS we can have
{\em non-determinism} in the sense that for two states $s_1$ $s_2$ we may have
$s_1 \ptrans{1:t_1} s_2$ and $s_1 \ptrans{1:t_2} s_2$, which would suggest
that it is possible to make a transition from $s_1$ to $s_2$ in different
times ($t_1$ and $t_2$) and probability $1$, i.e. certainly.
In order to eliminate non-determinism we will consider in this paper only
tPTS's which are subject to the following conditions:
\begin{enumerate}
\item for all $s \in S$ we have $\sum_{(s,p_i,t_j,s_k) \in \trans} p_i = 1$, and
\item for all $t \in T$ there is {\em at most one} tuple $(s_1,t,p,s_2) \in
  \trans$.
\end{enumerate}

The first condition means that we consider here a {\em purely probabilistic}
or {\em generative} execution model. The second condition allows us to
associate a unique probability to every transition time between two states,
i.e. triple $(s_1,t,s_2)$; this means that we can define a function $\pi: S
\times \T \times S \rightarrow [0,1]$ such that $s_1 \ptrans{p:t} s_2$ iff
$\pi(s_1,t,p_2) = p$. Note however, that it is still possible to have
differently timed transitions between states, i.e. it is possible to have
$(s_1,t_1,p_2,s_2) \in \trans$ and $(s_1,t_2,p_2,s_2) \in \trans$ with $t_1
\neq t_2$.

If for all $s_1,s_2 \in S$ there exists at most one $(s_1,t,p,s_2)\in \trans$,
we can also represent a timed Probabilistic Transition System with labelled
states as a quadruple $(S,\trans,\tau,\la)$ with $\tau: S \times S \rightarrow
[0,1] \times \T$, a timing function. Thus, to any two states $s_1$ and $s_2$
we associate a unique transition time $t_{s_1,s_2}$ and probability
$p_{s_1,s_2}$.

\begin{definition}
  Consider a tPTS $(S,\trans,\la)$ and an {\em initial
    state} $s_0 \in S$. An {\em execution sequence} or {\em trace} starting in
  $s_0$ is a sequence $(s_0,s_1,\ldots)$ such that
  $s_i\ptrans{p_i:t_i}s_{i+1}$, for all $i=0,1,2,\dots$.
\end{definition}

We associate, in the obvious way, to an execution sequence $\si =
(s_0,s_1,\ldots)$ three more sequences: (i) the transition probability
sequence: $(p_1,p_2,\ldots)$, (ii) a time stamp sequence: $(t_1,t_2,\ldots)$,
and (iii) a state label sequence: $(\la(s_o),\la(s_1),\ldots)$.

Even for a tPTS with a finite number of states it is possible to have infinite
execution sequences. It is thus, in general, necessary to consider measure
theoretic notions in order to define a mathematically sound model for the
possible behaviours of a tPTS.  However, as long as we consider only
terminating systems, i.e. finite traces, things are somewhat simpler. In
particular, in this case, probability distributions can replace measures as
they are equivalent.  In this paper we restrict our attention to terminating
traces and probability distributions. This allows us to define for every
finite execution sequence $\si = (s_0,s_1,\ldots)$ its {\em running time} as
$\tau(\si) = \sum t_i$, and its {\em execution probability} as $\pi(\si) =
\prod t_i$. We will also associate to every state $s_0$ its {\em execution
  tree}, i.e. the collection of all execution sequences starting in $s_0$.


\subsection{Observing tPTS's}

In Section~\ref{pWhile}\ we will present an operational semantics of a simple
imperative programming language, pWhile, via a tPTS. Based on this model we
will then investigate the vulnerability against attackers who are able to
observe (i) the time, and (ii) the state labels, i.e. the low variables. In
this setting we will argue that the combined observation of time and low
variables is more powerful than the observation of time and low variables
separately. The following example aims to illustrate this aspect which comes
from the properties of joint probability distributions.

\begin{example}
  In order to illustrate the role of joint distributions in the observation of
  timed PTS's let us consider the following simple systems.
  
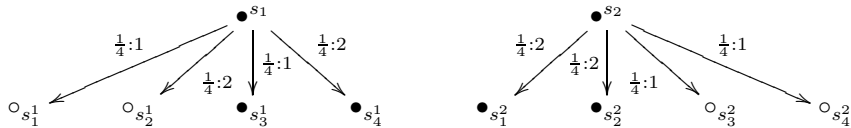
\[
  \xymatrix{
    &&
    \bullet^{s_1}
    \ar[dll]_{\frac{1}{4}:1}
    \ar[dl]^{\frac{1}{4}:2}
    \ar[d]^{\frac{1}{4}:1}
    \ar[dr]^{\frac{1}{4}:2}
    &
    \\
    \circ_{s^1_1}
    &
    \circ_{s^1_2} 
    &
    \bullet_{s^1_3} 
    &
    \bullet_{s^1_4} 
    \\
  }
  \hspace{1cm}
  \xymatrix{
    &
    \bullet^{s_2} 
    \ar[dl]_{\frac{1}{4}:2}
    \ar[d]_{\frac{1}{4}:2}
    \ar[dr]_{\frac{1}{4}:1}
    \ar[drr]^{\frac{1}{4}:1}
    &&
    \\
    \bullet_{s^2_1} 
    &
    \bullet_{s^2_2} 
    &
    \circ_{s^2_3} 
    &
    \circ_{s^2_4} 
    \\
  }
  \]
  We assume that the attacker can observe the execution times and that he/she
  is also able to (partially) distinguish (the final) states. In our example
  we assume that the states depicted as $\bullet$ and $\circ$ form two classes
  which the attacker can identify (e.g. because $\bullet$ and $\circ$ states
  have the same values for low, variables). The question now is whether this
  information allows the attacker to distinguish the two tPTS's.

  If we consider the information obtained by observing the running time, we
  see that both systems exhibit the same time behaviour corresponding to the
  distribution $\{ \tup{1,\frac{1}{2}}, \{\tup{2,\frac{1}{2}} \}$ over $\T =
  \{1,2\}$. The same is true in the case where the information is obtained by
  inspecting the final states: we have the distributions $\{
  \tup{\bullet,\frac{1}{2}}, \{\tup{\circ,\frac{1}{2}} \}$ over
  $\L=\{\bullet,\circ\}$ for both systems.

  However, considering that the attacker can observe running time and labels
  simultaneously, we see that the system on the right hand side always runs
  for $2$ time steps iff it ends up in a $\bullet$ state and $1$ time step iff
  it ends up in a $\circ$ state. In the system on the left hand side there is
  no such {\em correlation} between running time and final state. The
  difference between the two systems, which allows an attacker to distinguish
  them, is reflected in the joint distributions over $\T \times \L$. These can
  be expressed in matrix form for the two systems above as:
  \[
  \begin{array}{c|cc}
    \chi_1(t,l) & ~~~1~~~ & ~~~2~~~    \\ \hline
    \bullet& \frac{1}{4} & \frac{1}{4} \\
    \circ  & \frac{1}{4} & \frac{1}{4} \\ 
  \end{array}
  \hspace{2cm}
  \begin{array}{c|cc}
    \chi_2(t,l) & ~~~1~~~ & ~~~2~~~ \\ \hline
    \bullet& 0           & \frac{1}{2} \\
    \circ  & \frac{1}{2} & 0           \\ 
  \end{array}
  \]
  Note that while $\chi_1$ is the product of two {\em independent} probability
  distributions on $\T$ and $\L$ it is not possible to represent $\chi_2$ in
  the same way.
\end{example}


\section{An Imperative Language}
\label{pWhile}

We consider a language similar to that used in \cite{Agat00}\ with the
addition of a probabilistic choice construct. The syntax of the language is as
follows:
\[
\begin{array}{lrcl}
\mbox{Operators:}   & \op & \syndef & 
+ \synalt * \synalt - \synalt = \synalt != \synalt < \synalt <= \\
\mbox{Expressions:} & e & \syndef & 
v \synalt x \synalt e~\op~e \\
\mbox{Commands:}    & C, D & \syndef & 
x := e \synalt \SkipAsn ~x~e \synalt \ifS{(e)}{C}{D} \synalt \SkipIf ~e~C \\ 
& & & \synalt \whileS{(e)}{C} \synalt C;D \synalt \chooseL{p}{C}{D} \\
\mbox{Basic Values:}& v & \syndef & 
n \synalt \true \synalt \false
\end{array}
\]
The probabilistic choice is used in an essential way in the program
transformation presented later.
We also keep the language of types in \cite{Agat00}, although in a 
simplified form:
\[
\begin{array}{lll}
\mbox{Security levels} ~~~ & 
s ~\syndef~ \low \synalt \high & (\mbox{with } \low \leq \high 
                                 ~\mbox{and}~ s \leq s) \\
\mbox{Base types} & 
\overline{\tau} ~\syndef~ \Int \synalt \Bool \\
\mbox{Security types} & 
\tau~\syndef~\overline{\tau}_s
\end{array}
\]
and sub-typing relation:
\[
\frac{\displaystyle
s_1 \leq s_2}
{\displaystyle
\overline{\tau}_{s_1} \leq \overline{\tau}_{s_2}}.
\]

We will indicate by $E$ the state of a computation and denote by $E_L$ its
restriction to low variables, i.e. a state which is defined as $E$ for all the
low variables for which $E$ is defined, and is undefined otherwise.  We say
that two configurations $ \tup{E~|~C}$ and $\tup{E'~|~C'}$ are {\em low
  equivalent} if and only if $E_L=E'_L$ and we indicate this by $\tup{E~|~C}
=_L \tup{E'~|~C'}$.  In the following we will sometimes use for configurations
the shorthand notation $c, c_1, c_2, \dots, c', c'_1, \ldots$. We will also
denote by ${\Confs}$ the set of all configurations.


\subsection{SOS Semantics}

The operational semantics of pWhile -- except for the probabilistic choice
construct -- follows essentially the one presented in \cite{Agat00}. 
For the convenience of the reader we present here  all the
rules which are based on the big step semantics for expressions (where
$\sem{op}$ represents the usual semantics of operators):

\[
\begin{array}{l@{~~~~~~~~}c@{~~~~~~~~}r}
E \vdash v \Downarrow v 
&
\frac{\displaystyle E(x) = v}{\displaystyle E \vdash x \Downarrow v} 
&
\frac{
\displaystyle E \vdash e_1 \Downarrow v_1 \hspace{0.5cm} 
              E \vdash e_2 \Downarrow v_2}{
\displaystyle E \vdash e_1 ~\op~ e_2 \Downarrow v_1 \sem{op} v_2}
\end{array}
\]

The small step semantics is then define as a timed PTS via the 
SOS rules in Table~\ref{OpSem}.

\begin{table}[t]
\hrule
\[
\begin{array}{ll}
\mbox{(Assign)} & 
\frac{
\displaystyle 
E \vdash e \Downarrow v
}{
\displaystyle 
\tup{E ~|~ x:=e} \ptrans{1:t_e \cdot t_x \cdot t_{asn} \cdot \surd} E[x=v]
}
\\[5ex]
\mbox{(Seq)} &
\frac{
\displaystyle
\tup{E~|~C} \ptrans{p:ts \cdot \surd)} E'
}{
\displaystyle
\tup{E~|~C ; D} \ptrans{p:ts} \tup{E'~|~D}
}
\\[3ex] &
\frac{
\displaystyle
\tup{E~|~C} \ptrans{p:ts} \tup{E'~|~C'}
}{
\displaystyle
\tup{ E~|~C ; D} \ptrans{p:ts} \tup{E'~|~C' ; D}
}
\\[5ex]
\mbox{(If)} &
\frac{
\displaystyle
E \vdash e \Downarrow \true
}{
\displaystyle
\tup{E~|~\ifS{(e)}{C}{D}} \ptrans{1:t_e \cdot t_{br}}\tup{E~|~C}
}
\\[3ex] &
\frac{
\displaystyle
E \vdash e \Downarrow \false
}{
\displaystyle
\tup{E~|~\ifS{(e)}{C}{D}} \ptrans{1:t_e \cdot t_{br}} \tup{E~|~D}
}
\\[5ex]
\mbox{(SkipAsn)} &
\frac{
\displaystyle
E \vdash e \Downarrow v
}{
\displaystyle
\tup{E~|~\SkipAsn~ x~ e} \ptrans{1:t_e \cdot t_{asn} \cdot \surd} E
}
\\[5ex]
\mbox{(SkipIf)} &
\frac{
\displaystyle
E \vdash e \Downarrow v
}{
\displaystyle
\tup{E~|~\SkipIf~ e~ C} \ptrans{1:t_e \cdot t_{br}} \tup{E~|~C}
}~~~~~v \in \{\true,\false\}
\\[5ex]
\mbox{(While)} &
\frac{
\displaystyle
E \vdash e \Downarrow \false
}{
\displaystyle
\tup{E~|~\whileS{(e)}{C}} \ptrans{1:t_e \cdot t_{br} \cdot \surd} E
}
\\[3ex] &
\frac{
\displaystyle
E \vdash e \Downarrow \true
}{
\displaystyle
\tup{E~|~\whileS{(e)}{C}} \ptrans{1:t_e \cdot t_{br}} 
                           \tup{E~|~C ; \whileS{(e)}{C}} 
}
\\[5ex]
\mbox{(Choose)} &
\frac{
\displaystyle 
~
}{
\displaystyle
\tup{E~|~\chooseL{p}{C}{D}} \ptrans{p:t_{ch}} \tup{E~|~C}
}
\\[3ex] &
\frac{
\displaystyle 
~
}{
\displaystyle
\tup{E~|~\chooseL{p}{C}{D}} \ptrans{(1-p):t_{ch}} \tup{ E~|~D}
}
\end{array}
\]
\hrule\vspace{1em}
\caption{Operational Semantics}
\label{OpSem}
\end{table}

The time labels $t_.$ represent the time it takes to perform certain
operations: $t_x$ is the time to store a variable, $t_e$ is the time it takes
to evaluate an expression, $t_{asn}$ represents the time to perform an
assignment, $t_{br}$ is the time required for a branching step, and $t_{ch}$
is the time to perform a probabilistic choice. By $ts$ we denote any sequence
of time labels and with $\surd$ we indicate termination.

The rule (Choose) is the only new rule with respect to the original semantics
in \cite{Agat00}.  It states that the execution of a probabilistic choice
construct leads, after a time $t_{ch}$, to a state where either the command
$C$ or the command $D$ is executed with probability $p$ or $1-p$,
respectively.  This rule together with the standard transition rules for the
other constructs of the language define a tPTS for our pWhile language
according to Definition~\ref{tPTS}. In this tPTS, the state labels are given
by the environment, i.e. $\la(\tup{E~|~C}) = E$.


\subsection{Abstract Semantics}

According to the notion of security we consider in this paper, an observer or
attacker can only observe the changes in low variables. Therefore, we can
simplify the semantics by `collapsing' the execution tree in such a way that
execution steps during which the value of all low variables is unchanged are
combined into one single step.
We call an execution sequence $\si$ {\em deterministic} if $\pi(\si) = 1$, and
we call it {\em low stable} if $\la(s_i)|_L = l$ for all $s_i\in\si$. The empty
path (of length zero) is by definition deterministic and low stable.
An execution sequence is {\em maximal deterministic/low stable} if it is not a
proper sub-sequence of another deterministic/low stable path.

\begin{definition}
  We define the collapsed transition relation by: 
  $\tup{E_1~|~C_1} \pTrans{p:T} \tup{E_2~|~C_2}$ iff
\begin{description}
\item[(i)] there exists a configuration $\tup{E_1'~|~C_1'}$ such that
  $\tup{E_1~|~C_1} \ptrans{p:t} \tup{E_1'~|~C_1'}$,
\item[(ii)] the path $\tup{E_1'~|~C_1'} \ptrans{1:t_1} \ldots \ptrans{1:t_{n-1}}
  \tup{E_2'~|~C_2'} \ptrans{1:t_n} \tup{E_2~|~C_2}$ is deterministic,
\item[(iii)] the path $\tup{E_1~|~C_1} \ptrans{p:t} \tup{E_1'~|~C_1'}
  \ptrans{1:t_1} \ldots \ptrans{1:t_{n-1}} \tup{E_2'~|~C_2'}$ is maximal low
  stable,
\item[(iv)] and $T = t + {\displaystyle\sum_{i=1}^n t_i}$.
\end{description}
\end{definition}

This is illustrated in the following example. In the depicted execution trees
we indicate in the nodes only the state and omit the program parts of the
corresponding configurations. Moreover, we use the notation $[n,m]$ for the
state $E$ where $h$ has value $n$ and $l$ has value $m$. 
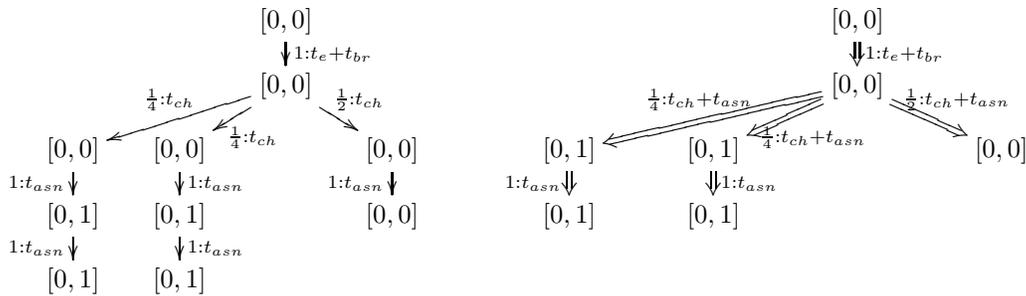
\[
  \xymatrix@R=3mm@C=5mm{
  &
  & 
  [0,0] 
  \ar[d]^{1:t_e + t_{br}}
  &
  \\
  &
  & 
  [0,0] 
  \ar[dll]_{\frac{1}{4}:t_{ch}}
  \ar[dl]^{\frac{1}{4}:t_{ch}}
  \ar[dr]^{\frac{1}{2}:t_{ch}}
  &
  \\
  [0,0] \ar[d]_{1:t_{asn}}
  &
  [0,0] \ar[d]^{1:t_{asn}}
  & 
  &
  [0,0] \ar[d]_{1:t_{asn}}
  \\
  [0,1] \ar[d]_{1:t_{asn}}
  &
  [0,1] \ar[d]^{1:t_{asn}}
  & 
  &
  [0,0] 
  \\
  [0,1]
  &
  [0,1]
  & 
  &
  }
~~~~~~~~
  \xymatrix@R=3mm@C=10mm{
  &
  & 
  [0,0] 
  \ar@{=>}[d]^{1:t_e + t_{br}}
  &
  \\
  &
  & 
  [0,0] 
  \ar@{=>}[dll]_{\frac{1}{4}:t_{ch} + t_{asn}}
  \ar@{=>}[dl]^{\frac{1}{4}:t_{ch} + t_{asn}}
  \ar@{=>}[dr]^{\frac{1}{2}:t_{ch} + t_{asn}}
  &
  \\
  [0,1] \ar@{=>}[d]_{1:t_{asn}}
  &
  [0,1] \ar@{=>}[d]^{1:t_{asn}}
  & 
  &
  [0,0] 
  \\
  [0,1] 
  &
  [0,1] 
  & 
  &
  }
\]
The collapsed execution tree on the right hand side represents in effect what
an attacker can actually observe during the program execution (for our
analysis of the situation we still record the value of $h$ although it is
invisible to the attacker).
 

\section{Bisimulation and Timing Leaks}

Observing the low variables and the running time separately is not the same as
observing them together; a correlation between the two random variables
(probability and time) has to be taken into account (cf.
Section~\ref{Model}).  A naive probabilistic extension of the
$\Gamma$-bisimulation notion introduced in \cite{Agat00} might not take this
into account. More precisely, this may happen if time and probability are
treated as two independent aspects which are observed separately in a mutual
exclusive way.  According to such a notion an attacker must set up two
different covert channels if she wants to exploit possible interference
through both the probabilistic and the timing behaviour of the system.

The notion of bisimulation we introduce here allows us to define a stronger
security condition: an attacker must be able to distinguish the probabilities
that two programs compute a given result in a given execution time. This is
obviously different from being able to distinguish the probability
distributions of the results {\em and} the running time.


\subsection{Probabilistic Time Bisimulation}
\label{PTBisim}

Probabilistic bisimulation was first introduced in \cite{LarsenSkou91}\ and
refers to an equivalence on probability distributions over the states of the
processes. This latter equivalence is defined as a lifting of the bisimulation
relation on the support sets of the distributions, namely the states
themselves.

An equivalence relation $\sim\;\subseteq S \times S$ on $S$ can be lifted to a
relation $\sim^* \;\subseteq \Dists(S) \times \Dists(S)$ between probability
distributions on $S$ via (cf \cite[Thm~1]{JonssonEtAl01}):
\[
\mu \sim^* \nu ~\mbox{iff}~ \forall [s] \in \Classes{S}{\sim}: \mu([s]) =
\nu([s]).
\]
It follows that $\sim^*$ is also an equivalence relation
(\cite[Thm~3]{JonssonEtAl01}).

For any equivalence relation $\sim$ on the set ${\Confs}$ of configurations,
we define the associated {\em low equivalence} relation $\sim_L$ by $c_1
\sim_L c_2$ if $c_1 \sim c_2$ and $c_1 =_L c_2$.  Obviously $\sim_L$ is again
an equivalence relation.  We can lift a low equivalence $\sim_L$ to
$(\sim_L)^*$ which we simply denote by $\sim_L^*$.

 
\begin{definition}
  Given a security typing $\GA$, a probabilistic time bisimilarity $\sim$ is
  the largest symmetric relation on configurations such that whenever
  $c_1 \sim c_2$,
  then
  \[
  c_1 \ttrans \chi_1
  ~\mbox{implies that there exists }~ 
  \chi_2 ~\mbox{ such that }~
  c_2 \ttrans \chi_2
  ~\mbox{and}~
  \chi_1 \sim_L^* \chi_2.
  \]

  We say that two configurations are probabilistic time bisimilar or
  PT-bisimilar, $c_1 \sim c_2$, if there exists a probabilistic time
  bisimilarity relation in which they are related.
\end{definition}

This definition generalises the one in \cite{Agat00}\ which only applies to
deterministic transition systems. Note that there is a difference between
$\sim_L^* = (\sim_L)^*$ and $(\sim^*)_L$; in fact, only the former is able to
take into account the correlation between time and low variables, while the
latter would be a straightforward generalisation of the time bisimulation in
\cite{Agat00} which is unable to model such a correlation.


We now exploit the notion of bisimilarity introduced above in order to
introduce a security property ensuring that a system is confined against any
combined attacks based on both timing and probabilistic covert channels.

\begin{definition}
  A pWhile program $P$ is {\em probabilistic time secure} or PT-secure if for
  any set of initial states $E$ and $E'$ such that $E_L = E'_L$, we have
  $\tup{E,P} \sim \tup{E',P}$.
\end{definition}


\section{Computing Approximate Bisimulation}

The papers \cite{TCS05,CONCUR03}\ introduce an approximate version of
bisimulation and confinement where the approximation can be used as a measure
$\varepsilon$ for the information leakage of the system under analysis. The
quantity $\varepsilon$ is formally defined in terms of the norm of a linear
operator representing the partition induced by the `minimal' bisimulation on
the set of the states of a given system, i.e. the one minimising the
observational difference between the system's components. We show here how to
compute a non-trivial upper bound $\delta$ to $\varepsilon$ by essentially
exploiting the algorithmic solution proposed by Paige and Tarjan
\cite{PaigeTarjan87}\ for computing bisimulation equivalence. This was already
adapted to PTS's in \cite{JLAP06}, where it was used for constructing a
padding algorithm as part of a transformational approach to the timing leaks
problem. In this approach the computational paths of a program are transformed
so as to make it perfectly secure by eliminating any possible timing covert
channel while preserving its I/O behaviour.

The algorithm we present here is an instantiation of that algorithm where the
abstract labels are replaced by the statements in a concrete language (pWhile)
and their execution times. Moreover, instead of transforming the execution
trees, our algorithm accumulates the information about the difference between
their transition probabilities and uses this information to compute an upper
bound $\delta$ to the maximal information leakage of the given program.


\subsection{Computing $\delta$ for PT-Bisimulation}


\begin{algorithm}
\begin{algorithmic}[1]
  \Procedure{QLumping}{$T_1,T_2$}
  \Assume $T_1$ execution tree with states $S_1$, and
          $T_2$ execution tree with states $S_2$
  \State $\delta \gets 0$
  \State $n \gets 0$
  \State $P \gets \{ S_1 \cup S_2 \}$    
    \Comment{(Initial) Partition}
  \While{$n \leq \Call{Height}{T_1 \oplus T_2}$}
     \State $S \gets \{ B \cap \Call{CutOff}{T_1 \oplus T_2,n}) ~|~ B \in P \}$
       \Comment{Splitters (below)}
     \While{$S \neq \emptyset$}
       \State choose $B\in S$, $S \gets S \setminus B$
         \Comment{Choose a splitter}
       \State $P \gets \Call{Splitting}{B,P}$
         \Comment{Split partition}
     \EndWhile
     \State $L_1 \gets \Call{Layer}{T1,n}$, 
            $L_2 \gets \Call{Layer}{T2,n}$
     \State $\Call{CompDelta}{L_1,L_2}$
     \State $n \gets n+1$
       \Comment{Go to next level}
  \EndWhile
         \State \Return $\delta$
  \EndProcedure
\end{algorithmic}
\caption{Algorithm for detecting critical blocks}
\label{QLumping}
\end{algorithm}


\begin{algorithm}[t]
\begin{algorithmic}[1]
  \Procedure{CompDelta}{$L_1,L_2$}
	\While{$L_1 \neq \emptyset$}
	  \State choose $s_1 \in L_1$, $L_1 \gets L_1 \setminus s_1$
	    \Comment For all $s_1\in L_1$
	  \State $\beta \gets \infty$
	  \State $L \gets L_2$	
	  \While{$L_2 \neq \emptyset$}
	    \State choose $s_2 \in L$, $L \gets L \setminus s_2$	 
	      \Comment For all $s_2\in L_2$
	    \State $\beta \gets \min(\beta,\|\chi(s_1)-\chi(s_2)\|_\infty)$
	      \Comment Find best match
	  \EndWhile
      \State $\delta \gets \max(\delta,\beta)$
    \EndWhile
  \EndProcedure
\end{algorithmic}
\caption{Algorithm for computing $\delta$}
\label{CompDelta}
\end{algorithm}


Algorithm~\ref{CompDelta}\ describes a procedure that can be used inside an
algorithm for constructing a lumping (i.e.\ a PT-bisimulation equivalence) of
two tPTS's $T_1$ and $T_2$. In particular, Algorithm~\ref{QLumping}\ refers to
a such a procedure which follows the algorithmic paradigm for partition
refinement introduced by Paige and Tarjan in \cite{PaigeTarjan87} (see also
\cite{DerisaviEtAl03,DovierEtAl04}). The Paige-Tarjan algorithm constructs a
partition of a state space $\Sigma$ which is {\em stable} for a given
transition relation $\rightarrow$. It is a well-known result that this
partition corresponds to a bisimulation equivalence on the transition system
$(\Sigma,\rightarrow)$. The refinement procedure used in the algorithm
consists in {\em splitting} the blocks in a given partition $P$ by replacing
each block $B\in P$ with $B\cap pre{S}$ and $B\setminus pre{S}$, where
$S\subseteq \Sigma$ and $pre(X)=\{s\in \Sigma\mid s\rightarrow x\mbox{ for
  some }x\in X\}$.

In order to check whether two execution trees $T_1$ and $T_2$ in our tPTS
model are PT-bisimilar, in Algorithm~\ref{QLumping} we apply this refinement
technique to the set of states formed by the disjoint union of the states in
$T_1$ and $T_2$.  The strategy of our lumping procedure $\Call{QLumping}{T_1,
  T_2}$ \ is as follows: it proceeds iteratively layer by layer starting from
the leaves layer, and splits the blocks in the current partition restricted to
the current layer.  The procedure $\Call{CompDelta}{L_1,L_2}$ computes for
each two layers $L_1$ and $L_2$, the maximal difference
$\|\chi(s_1)-\chi(s_2)\|_\infty$ between the probabilities to get from states
in $T_1\cap L_1$ and $T_2\cap L_1$, respectively, into states of layer $L_2$.
In the original lumping procedure this would determine a splitting of the
states in layer $L_1$. This value is stored in a variable $\beta$ and compared
with the current value of a variable $\delta$ which contains the maximal
difference up to that iteration. When the lumping algorithm terminates (that
is when we have reached the root of the union tree), one of the following
situations will occur: either the roots of $T_1$ and $T_2$ belong to the same
class in the constructed partition (i.e. $T_1$ and $T_2$ are PT-bisimilar) or
not. In the latter case $\delta$ will contain a maximal difference in the
transition probabilities of the two processes which makes them non-bisimilar.
This is therefore an estimate of the information leakage of the system. Note
that, by construction, $\delta$ will be zero in the first case.

The strategy for constructing the lumping described above determines the {\em
  coarsest partition} of a set which is stable wrt a given relation
\cite{DerisaviEtAl03,DovierEtAl04}, that is in our case the coarsest
PT-bisimulation equivalence. Obviously, this does not necessarily coincide
with the `minimal' one corresponding to the quantity $\varepsilon$ defined in
\cite{TCS05}. Thus, $\delta$ will be in general only a safe approximation,
namely an upper bound to the capacity of probabilistic timing covert channel
defined by $\varepsilon$.
The following proposition is therefore a corollary of Proposition~45 in
\cite{TCS05}\ stating a similar assertion for $\varepsilon$-bisimulation.

\begin{proposition}
  $P$ is PT-secure iff for any pair of initial configurations $c_1,c_2$ the
  corresponding execution trees $T_1$ and $T_2$ are such that
  $\Call{QLumping}{T_1, T_2}$ returns $\delta =0$.
\end{proposition}


\subsection{A Weighted Version:  $\delta'$}

The actual value of $\de$ is determined by the way we compute the best match
between the joint probability distributions $\chi(s_1)$ and $\chi(s_2)$ in
line~8 of $\Call{CompDelta}{L_1,L_2}$. In order to compute $\de$ we use the
{\em supremum norm}, $\|\cdot\|_\infty$, between two distributions, i.e. the
largest absolute difference between corresponding entries in $\chi(s_1)$ and
$\chi(s_2)$, respectively. In other words, we try to identify a class of
states $C$ (in the layer below) and a time interval $t$ such that the
probability of reaching this class in that time from $s_1$ differs maximally
from the one for $s_2$.

One can argue that this is a fair approach as we treat all classes and time
labels the same way. However, it might be useful to develop a measure which
reflects the fact that certain times and classes are more similar than others.

 From the point of view of the attacker, such a measure would encode her/his
ability in detecting similarity as given by the nature and the precision of
the instruments he is actually using.  For example, suppose it is possible to
reach the same class $C$ from $s_1$ and $s_2$ with different times $t_1$ and
$t_2$, such that the corresponding probabilities determine $\de$ (i.e. we have
the maximal difference in this case). However, we might in certain
circumstances also want to express the fact that $t_1$ and $t_2$ are more or
less similar, e.g. for $t_1=10$ and $t_2=10.5$ we might want a smaller $\de'$
than for $t_1=1$ and $t_2=100$.  In terms of the attacker, this means that we
make our estimate dependent on the actual power of the time detection
instrument that he/she possesses.

In order to incorporate similarity of times and/or classes we need to modify
the way we determine the best match in line~8 of $\Call{CompDelta}{L_1,L_2}$.
Instead of determining the norm between $\chi(s_1)$ and $\chi(s_2)$ we can
compute a weighted version as:
\[
\beta \gets 
\min(\beta,\|\omega\cdot\chi(s_1)-\omega\cdot\chi(s_2)\|_\infty) =
\min(\beta,\|\omega\cdot(\chi(s_1)-\chi\cdot(s_2))\|_\infty),
\]
where $\omega$ re-scales the entries in $\chi(s_1)$ and $\chi(s_2)$ so as to
reflect the relative importance of certain times and/or classes. Note that
``$\cdot$'' denotes here the component-wise and not the matrix multiplication:
$(\omega\cdot\chi)_{tC} = \omega_{tC}\chi_{tC}$. If, for example, an attacker
is not able to detect the absolute difference between times but can only
measure multiplicities expressing approximative proportions, we could re-scale
the $\chi$'s via $\omega_{tC} = \log(t)$.

In the following we will use a weighted version $\de'$ which reflects the
similarity of classes. The idea is to weight according to the
``replaceability'' of a class. To this purpose we associate to every class (in
the layers below) a matching measure $\mu(C) = \min_{C \neq C'}\de'(C,C')$,
i.e. we determine the $\de'$ between a (sub)tree with a root in the class $C$
in question and all (sub)trees with roots in any of the other classes $C'$.
We can take any representative of the classes $C$ and $C'$ as these are by
definition bisimilar. The measure $\mu$ indicates how easy it is to replace
class $C$ by another one, or how good/precise is the attacker in
distinguishing successor states.  Then $\de'$ is simply the weighted version
of $\de$ as described above with $\omega_{tC} = \mu(C)$. Note that there is no
problem with the fact that $\de'$ is defined recursively as we always know the
$\de'$ in the layers below before we compute $\de'$ in the current layer.


\begin{example}
  In order to illustrate how $\delta$ and $\delta'$ quantify the difference
  between various execution trees, let us consider the following four trees.
  
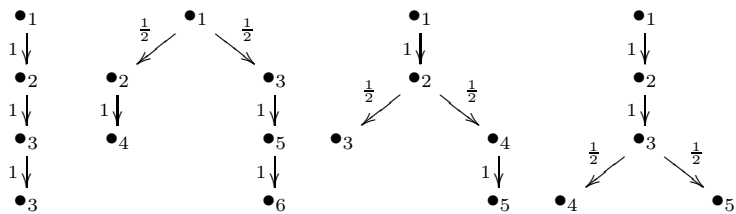
\[
  \xymatrix@R=4mm{
  \bullet_1 \ar[d]_1 \\
  \bullet_2 \ar[d]_1 \\
  \bullet_3 \ar[d]_1 \\
  \bullet_3 \\
  }
  ~~~~~
  \xymatrix@C=5mm@R=4mm{
  & 
  \bullet_1 \ar[dl]_{\frac{1}{2}} \ar[dr]^{\frac{1}{2}}
  & \\
  \bullet_2 \ar[d]_1 
  &&
  \bullet_3 \ar[d]_1   
  \\
  \bullet_4 
  &&
  \bullet_5 \ar[d]_1   
  \\
  &&
  \bullet_6  
  \\
  }
  ~~~  
  \xymatrix@C=5mm@R=4mm{
  & 
  \bullet_1 \ar[d]_1
  & \\
  & 
  \bullet_2 \ar[dl]_{\frac{1}{2}} \ar[dr]^{\frac{1}{2}}
  & \\
  \bullet_3 
  &&
  \bullet_4 \ar[d]_1   
  \\
  &&
  \bullet_5
  \\
  }
  ~~~  
  \xymatrix@C=5mm@R=4mm{
  & 
  \bullet_1 \ar[d]_1
  & \\
  & 
  \bullet_2 \ar[d]_1
  & \\
  & 
  \bullet_3 \ar[dl]_{\frac{1}{2}} \ar[dr]^{\frac{1}{2}}
  & \\
  \bullet_4 
  &&
  \bullet_5 
  \\
  }
  \]
  We abstract from the influence of different transition times and individual
  state labels, i.e. we assume that $t=1$ for all transitions and that all
  states are labelled with the same label.
  
  If we compute the $\de$ and $\de'$ values between all the pairs of systems
  we get the following results:
  \[
  \begin{array}{c|rrrr}
    \de   & \bT_1 & \bT_2 & \bT_3 & \bT_4 \\ \hline
    \bT_1 & 0.000 & 0.500 & 1.000 & 0.000 \\
    \bT_2 & 0.500 & 0.000 & 1.000 & 0.500 \\
    \bT_3 & 1.000 & 1.000 & 0.000 & 1.000 \\
    \bT_4 & 0.000 & 0.500 & 1.000 & 0.000 \\
  \end{array}
  ~~~~~~~
  \begin{array}{c|rrrr}
    \de'  & \bT_1 & \bT_2 & \bT_3 & \bT_4 \\ \hline
    \bT_1 & 0.000 & 0.250 & 0.125 & 0.000 \\
    \bT_2 & 0.250 & 0.000 & 0.125 & 0.250 \\
    \bT_3 & 0.125 & 0.125 & 0.000 & 0.125 \\
    \bT_4 & 0.000 & 0.250 & 0.125 & 0.000 \\
  \end{array}
  \]
  From this we see that $\de$ and $\de'$ are symmetric, i.e. the difference
  between two systems is symmetric; that every system is bisimilar with
  itself, i.e.  $\de=0=\de'$ (as we have an empty diagonal); and that the
  difference between two systems is between zero and one with values in
  between very well possible.
\end{example}


\section{Cost Analysis}

In a recent article on ``Software Bugtraps'' in {\em The Economist} the
authors report on some ongoing research at NIST on ``Software Assurance
Metrics and Tool Evaluation'' \cite{TheEconomist08}. They claim that ``{\em
  The purpose of the research is to get away from the feeling that `all
  software has bugs' and say `it will cost this much money to make software of
  this kind of quality'}''.  They then conclude: ``{\em Rather than trying to
  stamp out bugs altogether, in short, the future of "software that makes
  software better" may lie in working out where the pesticide can be most
  cost-effectively applied}''.

Our aim is to introduce ``cost factors'' in a similar way into computer
security. Instead of trying to achieve perfect security we will look at the
trade-off between costs of security counter measures -- such as increased
average running time -- and the improvement in terms of security, which we can
measure via the $\delta$ introduced above. Even in simple examples we are able
to exhibit interesting effects.


\subsection{Security Typing}

In \cite{Agat00}\ Agat introduces a program transformation to remove covert
timing channels ({\em timing leaks}) from programs written in a sequential
imperative programming language.  The language used is a language of security
types with two security levels that is based on earlier work by Volpano and
Smith \cite{VolpanoSmith98a,VolpanoSmith98b}.  Whilst Volpano and Smith
restrict the condition in both while-loops and if-commands to being of the
lowest security level, Agat allows the condition in an if-command to be high
security providing that an external observer cannot detect which branch was
taken. He shows that if a program is typeable in his system, then it is secure
against timing attacks.  This result depends critically on a notion of
bisimulation; an if-command with a high security condition is only typeable if
the two branches are bisimilar.  Agat's notion of bisimilarity is timing aware
and based on a notion of low-equivalence which ensures stepwise
non-interference.  He does not give an algorithm for bisimulation checking.

If a program fails to type, Agat presents a transformation system to remove
the timing leak.  The transformation pads the branches of if-commands with
high security conditions with dummy commands.  The objective of the padding is
that both branches end up with the same timing and thus become
indistinguishable by an external observer.  The transformation utilises the
concept of a {\em low-slice}: for a given command $C$, its low-slice $C_L$ has
the same syntactic structure as $C$ but only has assignments to low security
variables; all assignments to high security variables and branching on high
security conditions are replaced by skip commands of appropriate duration.
The transformation involves extending the branches in a high security
if-command by adding the low-slice from the other branch.  The effect of this
transformation is that the timing of the execution of both branches are the
same and equal to the sum of timing of the two branches in the untransformed
program.  Agat demonstrates that the transformation is semantically sound and
that transformed programs are secure (correctness).

In order to extend this system to our language, we only have to add a rule for
the {\bf choose} statement (essentially a straight forward extension of the
rule for {\bf if}). In detail, we present the typing rules in
Table~\ref{SecTypes}. Note that the rule (If$_{H}$) refers to the {\em
  semantic} notion of timed bisimilarity (as introduced in
Section~\ref{PTBisim}).

\begin{table}[t]
\hrule
\[
\begin{array}{ll}
\mbox{(Assign$_{H}$)} &
\frac{
\displaystyle
\Gamma \vdash_\leq e : \overline{\tau}_s ~~~~
\Gamma \vdash_= x : \overline{\tau}_{H} ~~~~ s \leq H
}{
\displaystyle
\Gamma \vdash x := e : \SkipAsn~x~e
}
\\[5ex]
\mbox{(Assign$_{L}$)} &
\frac{
\displaystyle
\Gamma \vdash_\leq e : \overline{\tau}_{L} ~~~
\Gamma \vdash_= x : \overline{\tau}_{L}
}{
\displaystyle
\Gamma \vdash x := e : x := e
}
\\[5ex]
\mbox{(Seq)} &
\frac{
\displaystyle
\Gamma \vdash C : C_{L} ~~~~ \Gamma \vdash D : D_{L}
}{
\displaystyle
\Gamma C ; D : C_{L} ; D_{L}
}
\\[5ex]
\mbox{(If$_{H}$)} &
\frac{
\displaystyle
\Gamma \vdash_\leq e : \Bool_{H} ~~~~
\Gamma \vdash C : C_{L} ~~~~ \Gamma \vdash D : D_{L}
}{
\displaystyle
\Gamma \vdash \ifS{(e)}{C}{D} : \SkipIf~e~C_{L}
} ~~~~ C_{L} \sim D_{L} 
\\[5ex]
\mbox{(If$_{L}$)} &
\frac{
\displaystyle
\Gamma \vdash_\leq e : \Bool_{L} ~~~~
\Gamma \vdash C : C_{L} ~~~~ \Gamma \vdash D : D_{L}
}{
\displaystyle
\Gamma \vdash \ifS{(e)}{C}{D} : \ifS{(e)}{C_{L}}{D_{L}}
}
\\[5ex]
\mbox{(While)} &
\frac{
\displaystyle
\Gamma \vdash_\leq e : \Bool_{L} ~~~~ \Gamma \vdash C : C_{L}
}{
\displaystyle
\Gamma \vdash \whileS{(e)}{C} : \whileS{(e)}{C_{L}}
}
\\[5ex]
\mbox{(Choose)} &
\frac{
\displaystyle
\Gamma \vdash C : C_{L} ~~~~ \Gamma \vdash D : D_{L}
}{
\displaystyle
\Gamma \vdash \chooseL{p}{C}{D} : \chooseL{p}{C_{L}}{D_{L}}
}
\\[5ex]
\mbox{(SkipAsn)} &
\frac{
\displaystyle 
~
}{
\displaystyle
\Gamma \vdash \SkipAsn~x~e : \SkipAsn~x~e
}
\\[5ex]
\mbox{(SkipIf)} &
\frac{
\displaystyle 
\Gamma \vdash C : C_{L}
}{
\displaystyle
\Gamma \vdash \SkipIf~e~C : \SkipIf~e~C_{L}
}
\end{array}
\]
\hrule\vspace{1em}
\caption{Security Typing Rules}
\label{SecTypes}
\end{table}


\subsection{Probabilistic Transformation}

We consider a probabilistic variant of Agat's language.  Probabilities
play an important role in the transformation.  Rather than just adding the
low slice from the other branch to each branch of a high security conditional,
we transform each branch to make a probabilistic choice between its
padded and untransformed variant.  This allows us to trade-off the
increased run-time of the padded program versus the vulnerability to attack
of the untransformed program.  The transformation described is just one
on a whole spectrum of probabilistic transformations -- at the other extreme
we could probabilistically decide whether or not to execute each 
command in the low slice.
All the formal transformation rules for probabilistic padding are the same as
in \cite{Agat00}. The only exception is the rule (If$_{H}$): Here we replace
-- provided certain typing conditions are fulfilled -- the branches of an {\bf
  if} statement not just by the correctly ``padded'' version as in
\cite{Agat00}; instead we introduce in every branch a choice such that the
secure replacement will be executed only with probability $p$ while with
probability $1-p$ the original code fragment will be executed.

In order to transform programs into secure versions we need to introduce an
auxiliary notion, namely the notion of global effect $ge(C)$ of commands. This
is used to identify (global) variables which might be changed when a command
$C$ is executed. Here is its formal definition:
\[
\begin{array}{rcl}
ge(x := e)                & = & \{x\}\\
ge(C_1;C_2)               & = & ge(C_1) ~\cup~ ge(C_2)\\
ge(\ifS{(e)}{C_1}{C_2})   & = & ge(C_1) ~\cup~ ge(C_2)\\
ge(\whileS{(e)}{C})       & = & ge(C)\\
ge(\chooseL{p}{C_1}{C_2}) & = & ge(C_1) ~\cup~ ge(C_2)\\
ge(\SkipAsn~x~e)          & = & \emptyset\\
ge(\SkipIf~e~C)           & = & ge(C).
\end{array}
\]

The judgments or transformation rules in Table~\ref{ProgTrans}\ are of the
general form:
\[
\Gamma \vdash C \hookrightarrow D ~|~ D_{L}
\]
which represents the fact that with a certain (security) typing $\Gamma$ we
can transform the statement $C$ into $D$ -- we also recorder as a side-product
the so-called {\em low slice} $D_{L}$ of $D$. 

\begin{table}[t]
\hrule
\[
\begin{array}{ll}
\mbox{(Assign$_{H}$)} &
\frac{
\displaystyle
\Gamma \vdash_\leq e : \overline{\tau}_s ~~~~
\Gamma \vdash_= x : \overline{\tau}_{H} ~~~~ s \leq H
}{
\displaystyle 
\Gamma \vdash x := e \hookrightarrow x := e ~|~ \SkipAsn~x~e
}
\\[5ex]
\mbox{(Assign$_{L}$)} &
\frac{
\displaystyle
\Gamma \vdash_\leq e : \overline{\tau}_{L} ~~~~
\Gamma \vdash_= x : \overline{\tau}_{L}
}{
\displaystyle 
\Gamma \vdash x := e \hookrightarrow x := e ~|~ x := e
}
\\[5ex]
\mbox{(Seq)} &
\frac{\displaystyle
\Gamma \vdash C_1 \hookrightarrow D_1 ~|~ D_{1L} ~~~~
\Gamma \vdash C_2 \hookrightarrow D_2 ~|~ D_{2L}
}{
\displaystyle
\Gamma \vdash C_1 ; C_2 \hookrightarrow D_1 ; D_2 ~|~ D_{1L} ; D_{2L}
}
\\[5ex]
\mbox{(If$_{H}$)} &
\frac{
\displaystyle
\Gamma \vdash_\leq e: \Bool_{H} ~~
\Gamma \vdash C_1 \hookrightarrow D_1 ~|~ D_{1L} ~~
\Gamma \vdash C_2 \hookrightarrow D_2 ~|~ D_{2L} ~~
ge(D_{1L}) = \emptyset ~~ ge(D_{2L}) = \emptyset
}{
\displaystyle
\Gamma \vdash \ifS{(e)}{C_1}{C_2} \hookrightarrow
\begin{array}{l}
{
\displaystyle 
{\bf if}~(e)~{\bf then}~(\chooseL{p}{D_1}{D_1 ; D_{2L}})~{\bf else}
} 
\\
{
\displaystyle 
(\chooseL{p}{D_2}{D_{1L} ; D_2}) ~|~ \SkipIf~e~(D_{1L} ; D_{2L})
}
\end{array}
}
\\[7ex]
\mbox{(If$_{L}$)} &
\frac{
\displaystyle
\Gamma \vdash_\leq e : \Bool_{L} ~~
\Gamma \vdash C_1 \hookrightarrow D_1 ~|~ D_{1L} ~~
\Gamma \vdash C_2 \hookrightarrow D_2 ~|~ D_{2L}
}{
\displaystyle
\Gamma \vdash \ifS{(e)}{C_1}{C_2} \hookrightarrow
\ifS{(e)}{D_1}{D_2} ~|~
\ifS{(e)}{D_{1L}}{D_{2L}}
}
\\[5ex]
\mbox{(While)} &
\frac{
\displaystyle
\Gamma \vdash_\leq e : \Bool_{L} ~~~~
\Gamma \vdash C \hookrightarrow D ~|~ D_{L}
}{
\displaystyle
\Gamma \vdash \whileS{(e)}{C} \hookrightarrow 
\whileS{(e)}{D} ~|~ \whileS{(e)}{D_{L}}
}
\\[5ex]
\mbox{(Choose)} &
\frac{
\displaystyle
\Gamma \vdash C_1 \hookrightarrow D_1 ~|~ D_{1L} ~~~~
\Gamma \vdash C_2 \hookrightarrow D_2 ~|~ D_{2L}
}{
\displaystyle
\Gamma \vdash \chooseL{p}{C_1}{C_2} \hookrightarrow
\chooseL{p}{D_1}{D_2} ~|~ \chooseL{p}{D_{1L}}{D_{2L}}
}
\\[5ex]
\mbox{(SkipAsn)} &
\frac{
\displaystyle 
~
}{
\displaystyle
\Gamma \vdash \SkipAsn~x~e \hookrightarrow 
\SkipAsn~x~e ~|~ \SkipAsn~x~e
}
\\[5ex]
\mbox{(SkipIf)} &
\frac{
\displaystyle
\Gamma \vdash C \hookrightarrow D ~|~ D_{L}
}{
\displaystyle
\Gamma \vdash \SkipIf~e~C \hookrightarrow
\SkipIf~e~D ~|~ \SkipIf~e~D_{L}}
\end{array}
\]
\hrule\vspace{1em}
\caption{Probabilistic Program Transformation}
\label{ProgTrans}
\end{table}


\subsection{An Example}

Our probabilistic version of Agat's padding algorithm allows us to obtain {\em
  partially} fixed programs. Depending on the parameter $p$ with which we
introduce empty low slices to obfuscate the timing leaks we can determine the
(average) execution time of the fixed program in comparison with the
improvement in security.

Agat presents in his paper \cite{Agat00}\ an example which itself is based on
Kocher's study \cite{Kocher96}\ of timing attacks against the RSA algorithm.
In order to illustrate our approach we simplify the example slightly: The
insecure program {\tt agat} we start with is depicted on the left side in
Table~\ref{AgatProgs}. The fully padded version Agat's algorithm produces,
{\tt fagat}, is on the right hand side of Table~\ref{AgatProgs}\ (to keep
things simple we omit Agat's empty statements like {\tt skipAsn s s}; as {\tt
  skip} as well as {\tt s:=s} can be used just to `spend time' without having
any real effect on the store we can use e.g. {\tt s:=s} in place of Agat's
{\tt skipAsn s s}). The program, {\tt pagat}, presented in the middle of
Table~\ref{AgatProgs}\ is the result of {\em probabilistic padding}: The
original program {\tt agat} is transformed in such a way that the compensating
statements, i.e. low slices, are executed only with probability $p$ while with
probability $q=1-p$ the original code is executed.  For $p=0$ we have the same
behaviour as the original program {\tt agat} while for $p=1$ this program
behaves in the same way as Agat's fully padded version {\tt fagat}.

\begin{table}[t]
\hrule\vspace{1em}
\begin{center}
\begin{minipage}[t]{3.8cm}
\begin{verbatim}
i := 1;
while i<=3 do
 if k[i]==1 then
   s := s;  
 else 
   skip;
 fi;
 i := i+1;
od;
\end{verbatim}
\end{minipage}
~
\begin{minipage}[t]{5cm}
\begin{verbatim}
i := 1;
while i<=3 do
 if k[i]==1 then
   choose p: s := s; skip 
   or     q: s := s 
   ro 
 else 
   choose p: skip
   or     q: s := s; skip
   ro
 fi;
 i := i+1;
od;
\end{verbatim}
\end{minipage}
~
\begin{minipage}[t]{3cm}
\begin{verbatim}
i := 1;
while i<=3 do
 if k[i]==1 then
   s := s; skip 
 else 
   s := s; skip
 fi; 
 i := i+1;
od;
\end{verbatim}
\end{minipage}
\end{center}
\hrule\vspace{1em}
\caption{Versions of Agat's Program: {\tt agat},{\tt pagat}, and {\tt fagat}}
\label{AgatProgs}
\end{table}

In our concrete experiments we used the following assumptions. The variable
${\tt i}$ can take values in $\{1,..,4\}$ while ${\tt k}$ is a three
dimensional array with values in $\{0,1\}$ -- nothing is concretely assumed
about ${\tt s}$. The variables ${\tt k}$, representing a {\em secret key}, and
${\tt s}$ have security typing $H$, while ${\tt i}$ is the only low variable
which can be observed by an attacker. We implemented this example using
(arbitrary) execution times: $t_{asn}= 3$ (assign time), $t_{br} = 2$
(test/branch time), and $t_{skip} = 1$ (skip time), and $t_{ch} = 0$ (choice
time).

The abstract semantics for the {\tt pagat} program -- which only records
choice points and the moments in time when the low variable changes its value
-- produces the following execution trees if we start with keys {\tt k=011}
and {\tt k=010}:
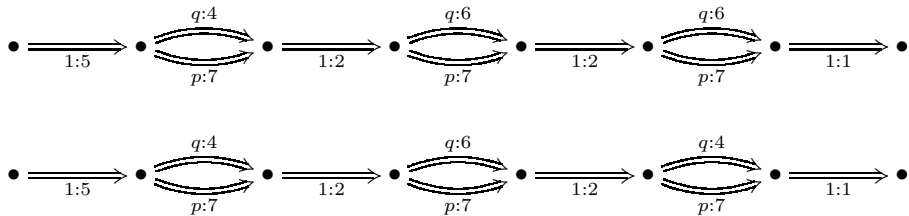
\[
  \xymatrix@C=13mm{
    \bullet \ar@2[r]_{1:5} &
    \bullet \ar@/^/@2[r]^{q:4} 
            \ar@/_/@2[r]_{p:7} &
    \bullet \ar@2[r]_{1:2} &
    \bullet \ar@/^/@2[r]^{q:6} 
            \ar@/_/@2[r]_{p:7} &
    \bullet \ar@2[r]_{1:2} &
    \bullet \ar@/^/@2[r]^{q:6} 
            \ar@/_/@2[r]_{p:7} &
    \bullet \ar@2[r]_{1:1} &
    \bullet
  }
\]
\[
  \xymatrix@C=13mm{
    \bullet \ar@2[r]_{1:5} &
    \bullet \ar@/^/@2[r]^{q:4} 
            \ar@/_/@2[r]_{p:7} &
    \bullet \ar@2[r]_{1:2} &
    \bullet \ar@/^/@2[r]^{q:6} 
            \ar@/_/@2[r]_{p:7} &
    \bullet \ar@2[r]_{1:2} &
    \bullet \ar@/^/@2[r]^{q:4} 
            \ar@/_/@2[r]_{p:7} &
    \bullet \ar@2[r]_{1:1} &
    \bullet
  }
\]
One can easily see from this how probabilistic padding influences the
behaviour of a program: For every bit in the key ${\tt k}$ -- i.e. every
iteration -- we have a choice between executing the original code with
probability $q=1-p$ or the `safe' code with probability $p$. The new code
always takes the same time (in our case $7$ ticks) while the original code's
execution time depends on whether ${\tt k[i]}$ is set or not (either $4$ or
$6$ time steps in our case). Clearly, for $p=0$ we get in every iteration a
different execution time, depending on the bit ${\tt k[i]}$, and thus can
deduce the secrete value ${\tt k}$ by just observing the execution times.
However, as the execution time is always the same for the replacement code, it
is impossible to do the same for $p=1$. For values of $p$ between $0$ and $1$,
the (average) execution times for ${\tt k[i]}=0$ and ${\tt k[i]}=1$ become
more and more similar. This means in practical terms that the attacker has to
spend more and more time (i.e. repeated observations of the program) in order
to determine with high confidence the exact execution time and thus deduce the
value of ${\tt k[i]}$ (cf. e.g. \cite{TCS05}).

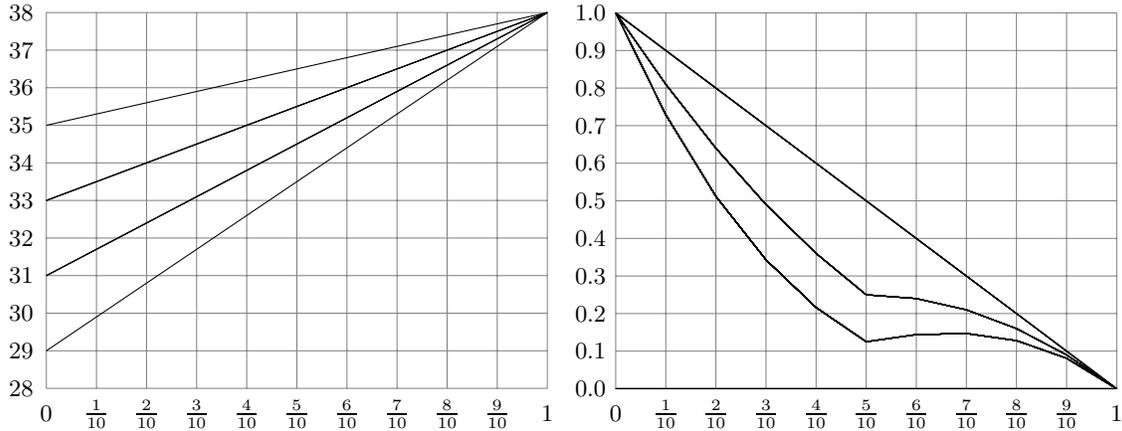
\begin{figure}[t]
\begin{tikzpicture}[y=5mm,x=6.66mm]
\draw[style=help lines] (0,28) grid[ystep=5mm,xstep=6.66mm] (10,38);
\draw (0,27.33) node{$0$};
\foreach \x in {1,...,9} {\draw (\x,27.33) node{$\frac{\x}{10}$};}
\draw (10,27.33) node{$1$};
\foreach \y in {28,...,38} {\draw (-0.5,\y) node{\y}; }
\draw (0.00,29.00) -- (1.00,29.90) -- (2.00,30.80) -- (3.00,31.70) --
(4.00,32.60) -- (5.00,33.50) -- (6.00,34.40) -- (7.00,35.30) -- (8.00,36.20)
-- (9.00,37.10) -- (10.00,38.00); \draw (0.00,31.00) -- (1.00,31.70) --
(2.00,32.40) -- (3.00,33.10) -- (4.00,33.80) -- (5.00,34.50) -- (6.00,35.20)
-- (7.00,35.90) -- (8.00,36.60) -- (9.00,37.30) -- (10.00,38.00); \draw
(0.00,31.00) -- (1.00,31.70) -- (2.00,32.40) -- (3.00,33.10) -- (4.00,33.80)
-- (5.00,34.50) -- (6.00,35.20) -- (7.00,35.90) -- (8.00,36.60) --
(9.00,37.30) -- (10.00,38.00); \draw (0.00,33.00) -- (1.00,33.50) --
(2.00,34.00) -- (3.00,34.50) -- (4.00,35.00) -- (5.00,35.50) -- (6.00,36.00)
-- (7.00,36.50) -- (8.00,37.00) -- (9.00,37.50) -- (10.00,38.00); \draw
(0.00,31.00) -- (1.00,31.70) -- (2.00,32.40) -- (3.00,33.10) -- (4.00,33.80)
-- (5.00,34.50) -- (6.00,35.20) -- (7.00,35.90) -- (8.00,36.60) --
(9.00,37.30) -- (10.00,38.00); \draw (0.00,33.00) -- (1.00,33.50) --
(2.00,34.00) -- (3.00,34.50) -- (4.00,35.00) -- (5.00,35.50) -- (6.00,36.00)
-- (7.00,36.50) -- (8.00,37.00) -- (9.00,37.50) -- (10.00,38.00); \draw
(0.00,33.00) -- (1.00,33.50) -- (2.00,34.00) -- (3.00,34.50) -- (4.00,35.00)
-- (5.00,35.50) -- (6.00,36.00) -- (7.00,36.50) -- (8.00,37.00) --
(9.00,37.50) -- (10.00,38.00); \draw (0.00,35.00) -- (1.00,35.30) --
(2.00,35.60) -- (3.00,35.90) -- (4.00,36.20) -- (5.00,36.50) -- (6.00,36.80)
-- (7.00,37.10) -- (8.00,37.40) -- (9.00,37.70) -- (10.00,38.00);
\end{tikzpicture}
\begin{tikzpicture}[y=5mm,x=6.66mm]
\draw[style=help lines] (0,0) grid[ystep=5mm,xstep=6.66mm] (10,10);
\draw (0,-0.66) node{$0$};
\foreach \x in {1,...,9} {\draw (\x,-0.66) node{$\frac{\x}{10}$};}
\draw (10,-0.66) node{$1$};
\foreach \y in {0,...,9} {\draw (-0.5,\y) node{0.\y}; }
\draw (-0.5,10) node{1.0};
\draw (0.00,0.00) -- (1.00,0.00) -- (2.00,0.00) -- (3.00,0.00) -- (4.00,0.00)
-- (5.00,0.00) -- (6.00,0.00) -- (7.00,0.00) -- (8.00,0.00) -- (9.00,0.00) --
(10.00,0.00); \draw (0.00,10.00) -- (1.00,7.29) -- (2.00,5.12) -- (3.00,3.43)
-- (4.00,2.16) -- (5.00,1.25) -- (6.00,1.44) -- (7.00,1.47) -- (8.00,1.28) --
(9.00,0.81) -- (10.00,0.00); \draw (0.00,10.00) -- (1.00,8.10) -- (2.00,6.40)
-- (3.00,4.90) -- (4.00,3.60) -- (5.00,2.50) -- (6.00,2.40) -- (7.00,2.10) --
(8.00,1.60) -- (9.00,0.90) -- (10.00,0.00); \draw (0.00,10.00) -- (1.00,7.29)
-- (2.00,5.12) -- (3.00,3.43) -- (4.00,2.16) -- (5.00,1.25) -- (6.00,1.44) --
(7.00,1.47) -- (8.00,1.28) -- (9.00,0.81) -- (10.00,0.00); \draw (0.00,10.00)
-- (1.00,9.00) -- (2.00,8.00) -- (3.00,7.00) -- (4.00,6.00) -- (5.00,5.00) --
(6.00,4.00) -- (7.00,3.00) -- (8.00,2.00) -- (9.00,1.00) -- (10.00,0.00);
\draw (0.00,10.00) -- (1.00,7.29) -- (2.00,5.12) -- (3.00,3.43) -- (4.00,2.16)
-- (5.00,1.25) -- (6.00,1.44) -- (7.00,1.47) -- (8.00,1.28) -- (9.00,0.81) --
(10.00,0.00); \draw (0.00,10.00) -- (1.00,8.10) -- (2.00,6.40) -- (3.00,4.90)
-- (4.00,3.60) -- (5.00,2.50) -- (6.00,2.40) -- (7.00,2.10) -- (8.00,1.60) --
(9.00,0.90) -- (10.00,0.00); \draw (0.00,10.00) -- (1.00,7.29) -- (2.00,5.12)
-- (3.00,3.43) -- (4.00,2.16) -- (5.00,1.25) -- (6.00,1.44) -- (7.00,1.47) --
(8.00,1.28) -- (9.00,0.81) -- (10.00,0.00); \draw (0.00,10.00) -- (1.00,7.29)
-- (2.00,5.12) -- (3.00,3.43) -- (4.00,2.16) -- (5.00,1.25) -- (6.00,1.44) --
(7.00,1.47) -- (8.00,1.28) -- (9.00,0.81) -- (10.00,0.00); \draw (0.00,0.00)
-- (1.00,0.00) -- (2.00,0.00) -- (3.00,0.00) -- (4.00,0.00) -- (5.00,0.00) --
(6.00,0.00) -- (7.00,0.00) -- (8.00,0.00) -- (9.00,0.00) -- (10.00,0.00);
\draw (0.00,10.00) -- (1.00,7.29) -- (2.00,5.12) -- (3.00,3.43) -- (4.00,2.16)
-- (5.00,1.25) -- (6.00,1.44) -- (7.00,1.47) -- (8.00,1.28) -- (9.00,0.81) --
(10.00,0.00); \draw (0.00,10.00) -- (1.00,8.10) -- (2.00,6.40) -- (3.00,4.90)
-- (4.00,3.60) -- (5.00,2.50) -- (6.00,2.40) -- (7.00,2.10) -- (8.00,1.60) --
(9.00,0.90) -- (10.00,0.00); \draw (0.00,10.00) -- (1.00,7.29) -- (2.00,5.12)
-- (3.00,3.43) -- (4.00,2.16) -- (5.00,1.25) -- (6.00,1.44) -- (7.00,1.47) --
(8.00,1.28) -- (9.00,0.81) -- (10.00,0.00); \draw (0.00,10.00) -- (1.00,9.00)
-- (2.00,8.00) -- (3.00,7.00) -- (4.00,6.00) -- (5.00,5.00) -- (6.00,4.00) --
(7.00,3.00) -- (8.00,2.00) -- (9.00,1.00) -- (10.00,0.00); \draw (0.00,10.00)
-- (1.00,7.29) -- (2.00,5.12) -- (3.00,3.43) -- (4.00,2.16) -- (5.00,1.25) --
(6.00,1.44) -- (7.00,1.47) -- (8.00,1.28) -- (9.00,0.81) -- (10.00,0.00);
\draw (0.00,10.00) -- (1.00,8.10) -- (2.00,6.40) -- (3.00,4.90) -- (4.00,3.60)
-- (5.00,2.50) -- (6.00,2.40) -- (7.00,2.10) -- (8.00,1.60) -- (9.00,0.90) --
(10.00,0.00); \draw (0.00,10.00) -- (1.00,8.10) -- (2.00,6.40) -- (3.00,4.90)
-- (4.00,3.60) -- (5.00,2.50) -- (6.00,2.40) -- (7.00,2.10) -- (8.00,1.60) --
(9.00,0.90) -- (10.00,0.00); \draw (0.00,10.00) -- (1.00,7.29) -- (2.00,5.12)
-- (3.00,3.43) -- (4.00,2.16) -- (5.00,1.25) -- (6.00,1.44) -- (7.00,1.47) --
(8.00,1.28) -- (9.00,0.81) -- (10.00,0.00); \draw (0.00,0.00) -- (1.00,0.00)
-- (2.00,0.00) -- (3.00,0.00) -- (4.00,0.00) -- (5.00,0.00) -- (6.00,0.00) --
(7.00,0.00) -- (8.00,0.00) -- (9.00,0.00) -- (10.00,0.00); \draw (0.00,10.00)
-- (1.00,7.29) -- (2.00,5.12) -- (3.00,3.43) -- (4.00,2.16) -- (5.00,1.25) --
(6.00,1.44) -- (7.00,1.47) -- (8.00,1.28) -- (9.00,0.81) -- (10.00,0.00);
\draw (0.00,10.00) -- (1.00,8.10) -- (2.00,6.40) -- (3.00,4.90) -- (4.00,3.60)
-- (5.00,2.50) -- (6.00,2.40) -- (7.00,2.10) -- (8.00,1.60) -- (9.00,0.90) --
(10.00,0.00); \draw (0.00,10.00) -- (1.00,7.29) -- (2.00,5.12) -- (3.00,3.43)
-- (4.00,2.16) -- (5.00,1.25) -- (6.00,1.44) -- (7.00,1.47) -- (8.00,1.28) --
(9.00,0.81) -- (10.00,0.00); \draw (0.00,10.00) -- (1.00,9.00) -- (2.00,8.00)
-- (3.00,7.00) -- (4.00,6.00) -- (5.00,5.00) -- (6.00,4.00) -- (7.00,3.00) --
(8.00,2.00) -- (9.00,1.00) -- (10.00,0.00); \draw (0.00,10.00) -- (1.00,7.29)
-- (2.00,5.12) -- (3.00,3.43) -- (4.00,2.16) -- (5.00,1.25) -- (6.00,1.44) --
(7.00,1.47) -- (8.00,1.28) -- (9.00,0.81) -- (10.00,0.00); \draw (0.00,10.00)
-- (1.00,7.29) -- (2.00,5.12) -- (3.00,3.43) -- (4.00,2.16) -- (5.00,1.25) --
(6.00,1.44) -- (7.00,1.47) -- (8.00,1.28) -- (9.00,0.81) -- (10.00,0.00);
\draw (0.00,10.00) -- (1.00,8.10) -- (2.00,6.40) -- (3.00,4.90) -- (4.00,3.60)
-- (5.00,2.50) -- (6.00,2.40) -- (7.00,2.10) -- (8.00,1.60) -- (9.00,0.90) --
(10.00,0.00); \draw (0.00,10.00) -- (1.00,7.29) -- (2.00,5.12) -- (3.00,3.43)
-- (4.00,2.16) -- (5.00,1.25) -- (6.00,1.44) -- (7.00,1.47) -- (8.00,1.28) --
(9.00,0.81) -- (10.00,0.00); \draw (0.00,0.00) -- (1.00,0.00) -- (2.00,0.00)
-- (3.00,0.00) -- (4.00,0.00) -- (5.00,0.00) -- (6.00,0.00) -- (7.00,0.00) --
(8.00,0.00) -- (9.00,0.00) -- (10.00,0.00); \draw (0.00,10.00) -- (1.00,7.29)
-- (2.00,5.12) -- (3.00,3.43) -- (4.00,2.16) -- (5.00,1.25) -- (6.00,1.44) --
(7.00,1.47) -- (8.00,1.28) -- (9.00,0.81) -- (10.00,0.00); \draw (0.00,10.00)
-- (1.00,8.10) -- (2.00,6.40) -- (3.00,4.90) -- (4.00,3.60) -- (5.00,2.50) --
(6.00,2.40) -- (7.00,2.10) -- (8.00,1.60) -- (9.00,0.90) -- (10.00,0.00);
\draw (0.00,10.00) -- (1.00,7.29) -- (2.00,5.12) -- (3.00,3.43) -- (4.00,2.16)
-- (5.00,1.25) -- (6.00,1.44) -- (7.00,1.47) -- (8.00,1.28) -- (9.00,0.81) --
(10.00,0.00); \draw (0.00,10.00) -- (1.00,9.00) -- (2.00,8.00) -- (3.00,7.00)
-- (4.00,6.00) -- (5.00,5.00) -- (6.00,4.00) -- (7.00,3.00) -- (8.00,2.00) --
(9.00,1.00) -- (10.00,0.00); \draw (0.00,10.00) -- (1.00,9.00) -- (2.00,8.00)
-- (3.00,7.00) -- (4.00,6.00) -- (5.00,5.00) -- (6.00,4.00) -- (7.00,3.00) --
(8.00,2.00) -- (9.00,1.00) -- (10.00,0.00); \draw (0.00,10.00) -- (1.00,7.29)
-- (2.00,5.12) -- (3.00,3.43) -- (4.00,2.16) -- (5.00,1.25) -- (6.00,1.44) --
(7.00,1.47) -- (8.00,1.28) -- (9.00,0.81) -- (10.00,0.00); \draw (0.00,10.00)
-- (1.00,8.10) -- (2.00,6.40) -- (3.00,4.90) -- (4.00,3.60) -- (5.00,2.50) --
(6.00,2.40) -- (7.00,2.10) -- (8.00,1.60) -- (9.00,0.90) -- (10.00,0.00);
\draw (0.00,10.00) -- (1.00,7.29) -- (2.00,5.12) -- (3.00,3.43) -- (4.00,2.16)
-- (5.00,1.25) -- (6.00,1.44) -- (7.00,1.47) -- (8.00,1.28) -- (9.00,0.81) --
(10.00,0.00); \draw (0.00,0.00) -- (1.00,0.00) -- (2.00,0.00) -- (3.00,0.00)
-- (4.00,0.00) -- (5.00,0.00) -- (6.00,0.00) -- (7.00,0.00) -- (8.00,0.00) --
(9.00,0.00) -- (10.00,0.00); \draw (0.00,10.00) -- (1.00,7.29) -- (2.00,5.12)
-- (3.00,3.43) -- (4.00,2.16) -- (5.00,1.25) -- (6.00,1.44) -- (7.00,1.47) --
(8.00,1.28) -- (9.00,0.81) -- (10.00,0.00); \draw (0.00,10.00) -- (1.00,8.10)
-- (2.00,6.40) -- (3.00,4.90) -- (4.00,3.60) -- (5.00,2.50) -- (6.00,2.40) --
(7.00,2.10) -- (8.00,1.60) -- (9.00,0.90) -- (10.00,0.00); \draw (0.00,10.00)
-- (1.00,7.29) -- (2.00,5.12) -- (3.00,3.43) -- (4.00,2.16) -- (5.00,1.25) --
(6.00,1.44) -- (7.00,1.47) -- (8.00,1.28) -- (9.00,0.81) -- (10.00,0.00);
\draw (0.00,10.00) -- (1.00,7.29) -- (2.00,5.12) -- (3.00,3.43) -- (4.00,2.16)
-- (5.00,1.25) -- (6.00,1.44) -- (7.00,1.47) -- (8.00,1.28) -- (9.00,0.81) --
(10.00,0.00); \draw (0.00,10.00) -- (1.00,9.00) -- (2.00,8.00) -- (3.00,7.00)
-- (4.00,6.00) -- (5.00,5.00) -- (6.00,4.00) -- (7.00,3.00) -- (8.00,2.00) --
(9.00,1.00) -- (10.00,0.00); \draw (0.00,10.00) -- (1.00,7.29) -- (2.00,5.12)
-- (3.00,3.43) -- (4.00,2.16) -- (5.00,1.25) -- (6.00,1.44) -- (7.00,1.47) --
(8.00,1.28) -- (9.00,0.81) -- (10.00,0.00); \draw (0.00,10.00) -- (1.00,8.10)
-- (2.00,6.40) -- (3.00,4.90) -- (4.00,3.60) -- (5.00,2.50) -- (6.00,2.40) --
(7.00,2.10) -- (8.00,1.60) -- (9.00,0.90) -- (10.00,0.00); \draw (0.00,10.00)
-- (1.00,7.29) -- (2.00,5.12) -- (3.00,3.43) -- (4.00,2.16) -- (5.00,1.25) --
(6.00,1.44) -- (7.00,1.47) -- (8.00,1.28) -- (9.00,0.81) -- (10.00,0.00);
\draw (0.00,0.00) -- (1.00,0.00) -- (2.00,0.00) -- (3.00,0.00) -- (4.00,0.00)
-- (5.00,0.00) -- (6.00,0.00) -- (7.00,0.00) -- (8.00,0.00) -- (9.00,0.00) --
(10.00,0.00); \draw (0.00,10.00) -- (1.00,7.29) -- (2.00,5.12) -- (3.00,3.43)
-- (4.00,2.16) -- (5.00,1.25) -- (6.00,1.44) -- (7.00,1.47) -- (8.00,1.28) --
(9.00,0.81) -- (10.00,0.00); \draw (0.00,10.00) -- (1.00,8.10) -- (2.00,6.40)
-- (3.00,4.90) -- (4.00,3.60) -- (5.00,2.50) -- (6.00,2.40) -- (7.00,2.10) --
(8.00,1.60) -- (9.00,0.90) -- (10.00,0.00); \draw (0.00,10.00) -- (1.00,8.10)
-- (2.00,6.40) -- (3.00,4.90) -- (4.00,3.60) -- (5.00,2.50) -- (6.00,2.40) --
(7.00,2.10) -- (8.00,1.60) -- (9.00,0.90) -- (10.00,0.00); \draw (0.00,10.00)
-- (1.00,7.29) -- (2.00,5.12) -- (3.00,3.43) -- (4.00,2.16) -- (5.00,1.25) --
(6.00,1.44) -- (7.00,1.47) -- (8.00,1.28) -- (9.00,0.81) -- (10.00,0.00);
\draw (0.00,10.00) -- (1.00,9.00) -- (2.00,8.00) -- (3.00,7.00) -- (4.00,6.00)
-- (5.00,5.00) -- (6.00,4.00) -- (7.00,3.00) -- (8.00,2.00) -- (9.00,1.00) --
(10.00,0.00); \draw (0.00,10.00) -- (1.00,7.29) -- (2.00,5.12) -- (3.00,3.43)
-- (4.00,2.16) -- (5.00,1.25) -- (6.00,1.44) -- (7.00,1.47) -- (8.00,1.28) --
(9.00,0.81) -- (10.00,0.00); \draw (0.00,10.00) -- (1.00,8.10) -- (2.00,6.40)
-- (3.00,4.90) -- (4.00,3.60) -- (5.00,2.50) -- (6.00,2.40) -- (7.00,2.10) --
(8.00,1.60) -- (9.00,0.90) -- (10.00,0.00); \draw (0.00,10.00) -- (1.00,7.29)
-- (2.00,5.12) -- (3.00,3.43) -- (4.00,2.16) -- (5.00,1.25) -- (6.00,1.44) --
(7.00,1.47) -- (8.00,1.28) -- (9.00,0.81) -- (10.00,0.00); \draw (0.00,0.00)
-- (1.00,0.00) -- (2.00,0.00) -- (3.00,0.00) -- (4.00,0.00) -- (5.00,0.00) --
(6.00,0.00) -- (7.00,0.00) -- (8.00,0.00) -- (9.00,0.00) -- (10.00,0.00);
\draw (0.00,10.00) -- (1.00,7.29) -- (2.00,5.12) -- (3.00,3.43) -- (4.00,2.16)
-- (5.00,1.25) -- (6.00,1.44) -- (7.00,1.47) -- (8.00,1.28) -- (9.00,0.81) --
(10.00,0.00); \draw (0.00,10.00) -- (1.00,7.29) -- (2.00,5.12) -- (3.00,3.43)
-- (4.00,2.16) -- (5.00,1.25) -- (6.00,1.44) -- (7.00,1.47) -- (8.00,1.28) --
(9.00,0.81) -- (10.00,0.00); \draw (0.00,10.00) -- (1.00,8.10) -- (2.00,6.40)
-- (3.00,4.90) -- (4.00,3.60) -- (5.00,2.50) -- (6.00,2.40) -- (7.00,2.10) --
(8.00,1.60) -- (9.00,0.90) -- (10.00,0.00); \draw (0.00,10.00) -- (1.00,7.29)
-- (2.00,5.12) -- (3.00,3.43) -- (4.00,2.16) -- (5.00,1.25) -- (6.00,1.44) --
(7.00,1.47) -- (8.00,1.28) -- (9.00,0.81) -- (10.00,0.00); \draw (0.00,10.00)
-- (1.00,9.00) -- (2.00,8.00) -- (3.00,7.00) -- (4.00,6.00) -- (5.00,5.00) --
(6.00,4.00) -- (7.00,3.00) -- (8.00,2.00) -- (9.00,1.00) -- (10.00,0.00);
\draw (0.00,10.00) -- (1.00,7.29) -- (2.00,5.12) -- (3.00,3.43) -- (4.00,2.16)
-- (5.00,1.25) -- (6.00,1.44) -- (7.00,1.47) -- (8.00,1.28) -- (9.00,0.81) --
(10.00,0.00); \draw (0.00,10.00) -- (1.00,8.10) -- (2.00,6.40) -- (3.00,4.90)
-- (4.00,3.60) -- (5.00,2.50) -- (6.00,2.40) -- (7.00,2.10) -- (8.00,1.60) --
(9.00,0.90) -- (10.00,0.00); \draw (0.00,10.00) -- (1.00,7.29) -- (2.00,5.12)
-- (3.00,3.43) -- (4.00,2.16) -- (5.00,1.25) -- (6.00,1.44) -- (7.00,1.47) --
(8.00,1.28) -- (9.00,0.81) -- (10.00,0.00); \draw (0.00,0.00) -- (1.00,0.00)
-- (2.00,0.00) -- (3.00,0.00) -- (4.00,0.00) -- (5.00,0.00) -- (6.00,0.00) --
(7.00,0.00) -- (8.00,0.00) -- (9.00,0.00) -- (10.00,0.00);
\end{tikzpicture}
\caption{Running Time $t(p)$ and Security Level $\de'(p)$ as Functions of $p$}
\label{AgatGraphs}
\end{figure}

The price we have to pay for increased security, i.e. indistinguishability of
behaviours, is an increased (average) execution time. The graph on the left in
Figure~\ref{AgatGraphs}\ shows how the running time (vertical axis) increases
in dependence of the padding probability $p$ (horizontal axis) for the eight
execution trees we have to consider in this example, i.e. for ${\tt k=000}$,
${\tt k=001}$, ${\tt k=010}$, etc. Depending on the number of bits set in
${\tt k}$ we get four different curves which show how, for example for ${\tt
  k=000}$ the running time increases from $29$ time steps (for $p=0$, i.e.
{\tt agat} program) to $38$ (for $p=1$, i.e. {\tt fagat} program).

We can employ the bisimilarity measures $\delta$ and $\delta'$ in order to
determine the security of the partially padded program. For this we compute
using our algorithm $\delta({\tt k}_i,{\tt k}_j)$ and $\delta'({\tt k}_i,{\tt
  k}_j)$ for all possible keys, i.e. $i,j=0,\ldots7$. It turns out that
$\delta=1$ for all values of $p<1$ and any pair of keys ${\tt k}_i$ and ${\tt
  k}_j$ with $i \neq i$; only for $p=1$ we get, as one would expect,
$\delta=0$ for all key pairs.  The weighted measure $\delta'$ is more
sensitive and we get for example for $p=0.5$ the following values when we
compare ${\tt k}_i$ and ${\tt k}_j$:
\[
\begin{array}{c|rrrrrrrr}
\delta' &{\tt000}&{\tt001}&{\tt010}&{\tt011}&{\tt100}&{\tt101}&{\tt110}&{\tt111}\\ 
\hline
{\tt000}& 0.000 & 0.125 & 0.250 & 0.125 & 0.500 & 0.125 & 0.250 & 0.125 \\
{\tt001}& 0.125 & 0.000 & 0.125 & 0.250 & 0.125 & 0.500 & 0.125 & 0.250 \\
{\tt010}& 0.250 & 0.125 & 0.000 & 0.125 & 0.250 & 0.125 & 0.500 & 0.125 \\
{\tt011}& 0.125 & 0.250 & 0.125 & 0.000 & 0.125 & 0.250 & 0.125 & 0.500 \\
{\tt100}& 0.500 & 0.125 & 0.250 & 0.125 & 0.000 & 0.125 & 0.250 & 0.125 \\
{\tt101}& 0.125 & 0.500 & 0.125 & 0.250 & 0.125 & 0.000 & 0.125 & 0.250 \\
{\tt110}& 0.250 & 0.125 & 0.500 & 0.125 & 0.250 & 0.125 & 0.000 & 0.125 \\
{\tt111}& 0.125 & 0.250 & 0.125 & 0.500 & 0.125 & 0.250 & 0.125 & 0.000 \\
\end{array}
\]
The diagonal entries are, of course, all zero as every execution tree is
bisimilar to itself. The other entries however are different from $0$ and $1$
and reflect the similarity between the two keys and thus the resulting
execution trees. If we plot the development of $\delta'$ as a function of $p$
we observe only three patterns as depicted in the right graph in
Figure~\ref{AgatGraphs}. In all three cases $\delta'$ decreases from an
original value $1$ to $0$, but in different ways.

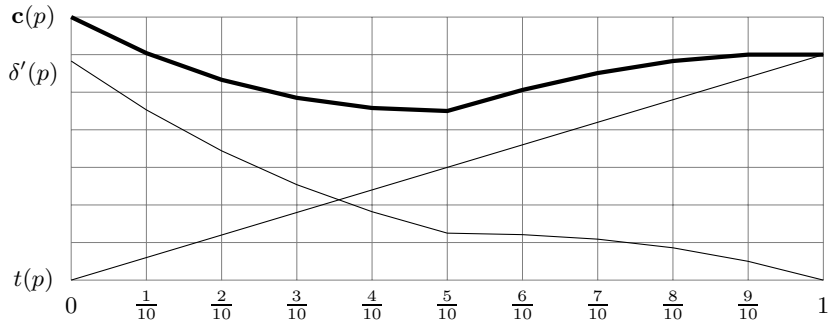
\begin{figure}[t]
\begin{center} 
\begin{tikzpicture}[y=5mm,x=10mm]
\draw[style=help lines] (0,3) grid[ystep=5mm,xstep=10mm] (10,10);
\draw (0,2.33) node{$0$};
\foreach \x in {1,...,9} {\draw (\x,2.33) node{$\frac{\x}{10}$};}
\draw (10,2.33) node{$1$};
\draw (-0.50,3.00) node{$t(p)$};
\draw (0.00,3.00) -- (1.00,3.60) -- (2.00,4.20) -- (3.00,4.80) -- (4.00,5.40)
-- (5.00,6.00) -- (6.00,6.60) -- (7.00,7.20) -- (8.00,7.80) -- (9.00,8.40) --
(10.00,9.00);
\draw (-0.50,8.50) node{$\de'(p)$};
\draw (0.00,8.83) -- (1.00,7.53) -- (2.00,6.44) -- (3.00,5.54) -- (4.00,4.82)
-- (5.00,4.25) -- (6.00,4.21) -- (7.00,4.09) -- (8.00,3.86) -- (9.00,3.50) --
(10.00,3.00);
\draw (-0.50,10.00) node{${\mathbf c(p)}$};
\draw[style=ultra thick] (0.00,10.00) -- (1.00,9.04) -- (2.00,8.33) --
(3.00,7.85) -- (4.00,7.58) -- (5.00,7.50) -- (6.00,8.06) -- (7.00,8.51) --
(8.00,8.83) -- (9.00,9.00) -- (10.00,9.00);
\end{tikzpicture}
\end{center}
\caption{Trade-Off and Costs $c(p)$ as a Functions of $p$}
\label{TradeOff}
\end{figure}

In analysing the trade-off between increased running time and security we need
to define a {\em cost} function. For example, one could be faced with a
situation where a certain code fragment needs to be executed in a certain
maximal time, i.e. there is a (cost) penalty if the execution takes longer
than a certain number of micro-seconds. In our case we will consider a very
simple cost function $c(p) = 6\delta'(p) + t(p)$ with $\delta'(p)$ and $t(p)$
the average $\delta'$ between all possible execution trees and $t$ the average
running time. The diagram in Figure~\ref{TradeOff}\ depicts how $c(p)$,
$\delta'(p)$ and $t(p)$ depend on the padding parameter $p$.

One can argue about the practical relevance of the particular cost function we
chose. Nevertheless, this example illustrates already nicely the non-linear
nature of security cost optimisation: The optimal, i.e. minimal, cost is
reached in this case for $p=0.5$, i.e. keeping the cost of security
counter measures in mind it is better to use a ``half-fixed'' program rather
than a completely safe one.


\section{Related and Further Work}

The idea of defining a secure system via the requirement that an attacker must
be unable to observe different behaviours as a result of different secrets --
i.e. the system ``operates in the same way'' whatever value a secret key has
-- goes back at least to the seminal work of Goguen and Meseguer
\cite{GoguenMeseguer82}.

This led in a number of settings to formalisations of security concepts such as
``non-interference'' via various notions of behavioural equivalencies (see
e.g. \cite{RyanSchneider99,FocardiGorrieri01}). One of the perhaps most
prominent of these equivalence notions, namely {\em bisimilarity}, plays an
important role in the context of security of concurrent systems but also found
application for sequential programs such as in Agat's work (as the interaction
between system and attacker can be modelled as a parallel composition).
In order to allow for a decision theoretic analysis of security
counter-measures and associated efforts it appears to be desirable to
introduce a ``quantitative'' notion of the underlying behavioural equivalence.
In the case of {\em bisimilarity} a first step was the introduction of the
notion of {\em probabilistic bisimulation} by Larson and Skou
\cite{LarsenSkou91}. However, this notion turns out to be still too strict and
a number of researchers developed ``approximate'' versions; among them we just
name the approaches by Desharnais et.al. \cite{Desharnais99,Desharnais02}\ and
van~Breugel \cite{vanBreugel05}\ and our work
\cite{CONCUR03,JCS04}\ (an extensive bibliography on this issue can be found in
\cite{ABE08}). We based this current paper on the latter approach because it allows for
an implementation of the semantics of pWhile via linear operators, i.e.
matrices, and an efficient computation of $\de$ and $\de'$ using standard
software such as {\tt octave} \cite{octave}.

Further research will be needed in order to clarify the relation between our
measures $\delta$ and existing notions of {\em approximate bisimilarity}
mentioned above, e.g. the $\ep$ in \cite{TCS05}. Furthermore, we also would
like to shed more light on the relationship between our notion and information
theoretic concepts used in the work of, for example Clark et.al.
\cite{ClarkEtAL05}\ and Boreale \cite{Boreale06}.



\newpage

\begin{thebibliography}{10}

\bibitem{VolpanoSmith98b}
Smith, G., Volpano, D.:
\newblock Secure information flow in a multi-threaded imperative language.
\newblock In: Proceedings of POPL'98, ACM Press (1998)  355--364

\bibitem{Kocher96}
Kocher, P.:
\newblock Timing attacks on implementations of {Diffie-Hellman}, {RSA}, {DSS},
  and other systems.
\newblock In: Proceedings of {CRYPTO} '96. Volume 1109 of Lecture Notes in
  Computer Science., Springer Verlag (1996)  104--113

\bibitem{Agat00}
Agat, J.:
\newblock Transforming out timing leaks.
\newblock In: Proceedings of POPL'00, ACM Press (2000)  40--53

\bibitem{alur94}
Alur, R., Dill, D.L.:
\newblock A theory of timed automata.
\newblock Theoretical Computer Science \textbf{126}(2) (1994)  183--235

\bibitem{KNSW04}
Kwiatkowska, M., Norman, G., Sproston, J., Wang, F.:
\newblock Symbolic model checking for probabilistic timed automata.
\newblock In Lakhnech, Y., Yovine, S., eds.: Proceedings of FORMATS/FTRTFT'04.
  Volume 3253 of Lecture Notes in Computer Science., Springer Verlag (2004)
  293--308

\bibitem{JonssonEtAl01}
Jonsson, B., Yi, W., Larsen, K.
\newblock In: Probabilistic Extentions of Process Algebras. Elsevier Science,
  Amsterdam (2001)  685--710

\bibitem{Stirzaker99}
Stirzaker, D.:
\newblock Probability and Random Variables.
\newblock Cambridge University Press (1999)

\bibitem{LarsenSkou91}
Larsen, K., Skou, A.:
\newblock Bisimulation through probabilistic testing.
\newblock Information and Computation \textbf{94} (1991)  1--28

\bibitem{TCS05}
{Di Pierro}, A., Hankin, C., Wiklicky, H.:
\newblock Measuring the confinement of probabilistic systems.
\newblock Theoretical Computer Science \textbf{340}(1) (2005)  3--56

\bibitem{CONCUR03}
{Di Pierro}, A., Hankin, C., Wiklicky, H.:
\newblock Quantitative relations and approximate process equivalences.
\newblock In Lugiez, D., ed.: Proceedings of CONCUR'03. Volume 2761 of Lecture
  Notes in Computer Science., Springer Verlag (2003)  508--522

\bibitem{PaigeTarjan87}
Paige, R., Tarjan, R.:
\newblock Three partition refinement algorithms.
\newblock SIAM Journal of Computation \textbf{16}(6) (1987)  973--989

\bibitem{JLAP06}
{Di Pierro}, A., Hankin, C., Siveroni, I., Wiklicky, H.:
\newblock Tempus fugit: How to plug it.
\newblock Journal of Logic and Algebraic Programming \textbf{72}(2) (2007)
  173--190

\bibitem{DerisaviEtAl03}
Derisavi, S., Hermanns, H., Sanders, W.H.:
\newblock Optimal state-space lumping in {M}arkov chains.
\newblock Information Processing Letters \textbf{87}(6) (September 2003)
  309--315

\bibitem{DovierEtAl04}
Dovier, A., Piazza, C., Policriti, A.:
\newblock An efficient algorithm for computing bisimulation equivalence.
\newblock Theoretical Computer Science \textbf{311}(1-3) (2004)  221--256

\bibitem{TheEconomist08}
{Software Bugtraps}:
\newblock Software that makes software better.
\newblock Economist \textbf{386}(8570) (March 2008)

\bibitem{VolpanoSmith98a}
Volpano, D., Smith, G.:
\newblock Confinement properties for programming languages.
\newblock SIGACT News \textbf{29}(3) (September 1998)  33--42

\bibitem{GoguenMeseguer82}
Goguen, J., Meseguer, J.:
\newblock Security {P}olicies and {S}ecurity {M}odels.
\newblock In: IEEE Symposium on Security and Privacy, IEEE Computer Society
  Press (1982)  11--20

\bibitem{RyanSchneider99}
Ryan, P., Schneider, S.:
\newblock Process algebra and non-interference.
\newblock Journal of Computer Security \textbf{9}(1/2) (2001)  75--103 Special
  Issue on CSFW-12.

\bibitem{FocardiGorrieri01}
Focardi, R., Gorrieri, R.:
\newblock Classification of {S}ecurity {P}roperties ({P}art {I}: {I}nformation
  {F}low).
\newblock In: Foundations of Security Analysis and Design - Tutorial Lectures.
  Volume 2171 of Lecture Notes in Computer Science., Springer Verlag (2001)
  331--396

\bibitem{Desharnais99}
Desharnais, J., Jagadeesan, R., Gupta, V., Panangaden, P.:
\newblock Metrics for labeled markov systems.
\newblock In: Proceedings of CONCUR'99. Volume 1664 of Lecture Notes in
  Computer Science., Springer Verlag (1999)  258--273

\bibitem{Desharnais02}
Desharnais, J., Jagadeesan, R., Gupta, V., Panangaden, P.:
\newblock The metric analogue of weak bisimulation for probabilistic processes.
\newblock In: Proceedings of LICS'02, IEEE (2002)  413--422

\bibitem{vanBreugel05}
{van Breugel}, F.:
\newblock A behavioural pseudometric for metric labelled transition systems.
\newblock In Abadi, M., de~Alfaro, L., eds.: Proceedings of CONCUR'05. Volume
  3653 of Lecture Notes in Computer Science., Springer Verlag (2005)  141--155

\bibitem{ABE08}
ABE'08:
\newblock Workshop on {A}pproximate {B}ehavioural {E}quivalences (2008) {\tt
  www.cse.yorku.ca/abe08}.

\bibitem{JCS04}
{Di Pierro}, A., Hankin, C., Wiklicky, H.:
\newblock Approximate {N}on-{I}nterference.
\newblock Journal of Computer Security \textbf{12}(1) (2004)  37--81

\bibitem{octave}
Eaton, J.W.:
\newblock Octave.
\newblock Technical report, Free Software Foundation, Boston, MA

\bibitem{ClarkEtAL05}
Clark, D., Hunt, S., Malacaria, P.:
\newblock Quantitative information flow, relations and polymorphic types.
\newblock Journal of Logic and Computation \textbf{15}(2) (2005)  181--199

\bibitem{Boreale06}
Boreale, M.:
\newblock Quantifying information leakage in process calculi.
\newblock In Bugliesi, M., Preneel, B., Sassone, V., Wegener, I., eds.:
  Proceedings of ICALP'06. Volume 4052 of Lecture Notes in Computer Science.,
  Springer Verlag (2006)  119--131

\end{thebibliography}
\end{document}